\documentclass[]{aa}   %

\usepackage{txfonts}
\usepackage{graphicx}
\usepackage{natbib}   \bibpunct{(}{)}{;}{a}{}{,}   %

\usepackage[bookmarksnumbered=true,breaklinks=true]{hyperref}

\begin{document}   %

\title{%
Are dust shell models well-suited to explain interferometric data of
late-type stars in the near-infrared?
}
\titlerunning{Dust shell models and interferometric data of late-type stars
in the near-infrared}

\author{%
P. Schuller\inst{1}\fnmsep\inst{2}
\and P. Salom\'e\inst{3} 
\and G. Perrin\inst{4} 
\and B. Mennesson\inst{5}
\and G. Niccolini\inst{6} \and P. de Laverny\inst{6}
\and S.T. Ridgway\inst{7}\fnmsep \inst{4} 
\and \\ 
	V. Coud\'e du Foresto\inst{4} 
\and W.A. Traub\inst{2}
}
\institute{%
Max-Planck-Institut f\"ur Astronomie, K\"onigstuhl 17, 69117 Heidelberg, Germany
\and 
Harvard-Smithsonian Center for Astrophysics, MS-20, 60 Garden Street, Cambridge, MA 02138, USA
\and 
Observatoire de Paris, LERMA, 61 Avenue de l'Observatoire, 75014 Paris, France
\and 
Observatoire de Paris, LESIA, 5 Place Jules Janssen, 92195 Meudon, France
\and 
Jet Propulsion Laboratory, MS 306-388, 4800 Oak Grove Drive, Pasadena, CA 91109, USA
\and 
Observatoire de la C\^{o}te d'Azur, D{\'e}partement Fresnel UMR 6528, BP 4229, 06304 Nice, France
\and 
National Optical Astronomy Observatories, P.O. Box 26732, Tucson, AZ 85726, USA
}
\offprints{Peter Schuller, \email{pschuller@cfa.harvard.edu}}

\date{Received ... / accepted ...}

\abstract{%
Recently available near-infrared interferometric data on late-type stars 
show a strong increase of diameter for asymptotic giant branch (AGB) stars 
between the K ($2.0 - 2.4~\mathrm{\mu m}$) and L ($3.4 - 4.1~\mathrm{\mu m}$) bands. 
Aiming at an explanation of these findings, 
we chose the objects 
\object{$\alpha$~Orionis} (\object{Betelgeuse}), \object{SW~Virginis}, and \object{R~Leonis}, 
which are of different spectral types and stages of evolution, 
and which are surrounded by circumstellar envelopes 
with different optical thicknesses. 
For these stars, we compared observations with spherically symmetric dust shell models. 
Photometric and $11~\mathrm{\mu m}$ interferometric data 
were also taken into account to further constrain the models. -- 
We find the following results. 
For all three AGB stars, the photosphere and 
dust shell model is consistent with the multi-wavelength photometric data.  
For $\alpha$~Orionis the model dust shell has a very small optical depth (0.0065 at $11~\mathrm{\mu m}$); 
the visibility data and model in K and L are essentially entirely 
photospheric with no significant contribution from the dust, 
and the visibility data at $11~\mathrm{\mu m}$ show a strong dust signature 
which agrees with the model. 
For SW~Virginis the model dust shell has a small optical depth (0.045 at $11~\mathrm{\mu m}$); 
in K the visibility data and model are essentially purely photospheric, 
in L the visibility data demand a larger object than the photosphere plus dust 
model allows, 
and at $11~\mathrm{\mu m}$ there was no data available. 
For R~Leonis the model dust shell has a moderate optical depth (0.1 at $11~\mathrm{\mu m}$); 
in K and L the visibility data and model situation is similar to that of SW~Vir, 
and at $11~\mathrm{\mu m}$ the visibility data and model are in agreement. -- 
We conclude that AGB models comprising a photosphere and dust shell, 
although consistent with SED data and also interferometric data in K and at $11~\mathrm{\mu m}$, 
cannot explain the visibility data in L; 
an additional source of model opacity, possibly related to a gas component, 
is needed in L to be consistent with the visibility data.

}   %

\maketitle   %

\section{Introduction}\label{sect.intro}
Supergiant stars and stars on the Asymptotic Giant Branch (AGB) are
surrounded by a dust envelope coupled with gas; 
for example, see \cite{cit.habing1996} for a review of these objects, 
\cite{cit.willson2000} for a review of mass loss, 
and \cite{cit.tsuji2000} for evidence of H$_2$O 
shells around supergiants. 

AGB stars are subject to pulsation and other surface phenomena 
(shock waves, acoustic waves etc., see \cite{cit.lafon1991}) 
that levitate matter above the stellar photosphere. 
Dust grains appear in the envelope at altitudes 
where the temperature falls below the condensation temperature. 
The radiation pressure drives the dust grain overflow far from the star 
up to distances of a thousand stellar radii causing mass-loss. 

The mechanism for mass loss in supergiants is not so clear. 
The main difference compared to Mira stars is that 
due to the larger luminosity of supergiants 
the atmosphere is much more extended 
and has a much greater pressure scale height. 
Therefore, the mass loss may have a different driving force 
although pulsation may also play a role. 

Both kinds of stars are characterised 
by their large diameter and strong luminosity 
and are consequently very good candidates for infrared interferometry. 
In the next section, we briefly present the instruments 
which were used to acquire the interferometric data discussed in this paper, 
namely FLUOR in the K band ($2.0 - 2.4~\mathrm{\mu m}$) 
and TISIS in the L band ($3.4 - 4.1~\mathrm{\mu m}$). 
We selected three particular stars of three different types 
for which we have a substantial set of data: 
the supergiant \object{$\alpha$~Ori}, the semi-regular \object{SW~Vir}, and the Mira-type star \object{R~Leo}. 
Furthermore, their envelopes have quite differing optical thicknesses. 
It appears that for the objects with large optical depth (SW~Vir and R~Leo) 
the observed diameters in L are considerably larger than 
the respective K band diameters 
(see Sect.~\ref{sect.ud-diam}). 
Measurements in the L band are more sensitive
to cooler material above the stellar 
photosphere than are the K band data. 
The question is: what is the physical nature of these layers? 
One possibility might be astronomical dust. 
Dust reprocesses the star radiation, in particular in the infrared. 
Depending on the dust abundance, 
the star's L band radiation might be superposed by a dust contribution 
that probes the inner region of the dust envelope 
at a few stellar radii away from the star. 
The measured L band radius might consequently be increased 
compared to the K band, which is mainly fed by stellar emission. 

In Sect.~\ref{sect.model}, the modelling of the
dust envelopes around these stars by a radiative transfer code is
explained and the parameters of the model are described.  
In Sect.~\ref{sect.comp}, we compare the model output 
with the interferometric and photometric data. 
Finally, in Sect.~\ref{sect.discuss} 
we discuss the limits of such models 
and the questions they raise about the L band observations.

\section{Observations and data reduction}\label{sect.obs-datared}
\subsection{Instruments used}\label{sect.instrumasp}
The stars discussed here were observed with the IOTA 
(Infrared-Optical Telescope Array) interferometer 
operating (at that time) with two telescopes \citep{cit.traub1998}. 
IOTA is located at the Smithsonian Institution's Whipple Observatory on Mount
Hopkins, Arizona. 
It offers multiple baseline observations 
providing visibilities at different spatial frequencies. 
This enables a model-fit to the visibility data to obtain information 
on the spatial intensity distribution of the sources. 
The beams were combined with FLUOR 
(Fiber Linked Unit for Optical Recombination) in the K band \citep{cit.foresto1998} 
and with its extension to the L band TISIS 
(Thermal Infrared Stellar Interferometric Set-up, \cite{cit.mennesson1999}). 
Beam combination with FLUOR is achieved by a single-mode fluoride glass 
triple coupler. 
The fibres filter the wavefronts corrugated by the atmospheric turbulence. 
The phase fluctuations are traded against photometric fluctuations 
which are monitored for each beam to correct for them a posteriori. 
The accuracy on visibility estimates measured by FLUOR is usually better 
than 1\% for most sources \citep{cit.teff} and can be as good 
as 0.2\% \citep{cit.perrin2003a}. 

TISIS is the extension of FLUOR 
to the thermal infrared ($3.4 - 4.1~\mathrm{\mu m}$). 
A single coupler is used as beam combiner and photometric signals are not monitored. 
Approximate correction of turbulence-induced flux fluctuations 
is achieved with the low-frequency part of the interferometric signal. 
Accuracy in the L band 
is therefore not as good as in the K band, but still better than 
without spatial filtering by fibres. 
The acquisition protocol is also 
different from that of the K band, 
since ground-based interferometric 
observations encounter a new difficulty at $3.6~\mathrm{\mu m}$. 
Thermal background emission produces a fluctuating photometric offset 
which must be correctly subtracted in order to achieve a good calibration. 
Hence each source observation must be bracketed 
by a sky background measurement (chopping) 
that reduces the instrument efficiency.

\subsection{Data selection}\label{sect.datasel}
We chose to include three stars in our study which span different
spectral types and stages of evolution, 
namely the supergiant $\alpha$~Orionis (Betelgeuse), 
the semi-regular variable SW~Virginis, and the Mira star R~Leonis. 
Some basic parameters of these objects are compiled in Table~\ref{Starscaract}. 
	\begin{table}
	\centering
	\caption{\label{Starscaract}
Some basic parameters of the studied objects taken from the General Catalogue of Variable Stars \citep{cit.gcvs1998}. 
	The distances come from the HIPPARCOS catalogue \citep{cit.hip1997}.}
	\begin{tabular}{ccccc}
	\hline\hline
Star  &  Spectral  &  Variability  &  Period  &  Distance  \\
	&  Type  &  	Type	&  (days)  &  (pc)  \\
	\hline
$\alpha$~Ori  &  M1I  &  SRc  &  2335  &  131  \\
SW~Vir        &  M7III  &  SRb  &  150  &  143  \\
R~Leo  	      &  M8IIIe  &  Mira  &  310  &  101  \\
	\hline
	\end{tabular}
	\end{table}
These stars are also some of the most observed in our sample. 
The L band data are extracted from \cite{cit.chagnon2002} -- 
we have used the data from the February-March 2000 observations. 
The K band data for $\alpha$~Orionis were presented in 
\cite{cit.perrin2002aori} and observations cover the period February~1996 
to March~1997. 
The K band data for SW~Virginis were collected in May~2000 
and are presented in \cite{cit.perrin2003a}. 
Lastly, the K band data for R~Leonis were published in \cite{cit.perrin1999} 
and were collected in April~1996 and March~1997. 
Except for SW~Virginis, 
there is a large time difference between the dates 
of the L band and K band data acquisitions. 
We believe this to not be a major issue - 
although these stars are known to be variable, 
the effects we want to analyse are 
slower than the periodic changes of stars and larger in amplitude. 

In order to get the best constraints on the dust shell model spatial 
distribution, we have included interferometric data acquired with ISI 
(Infrared Spatial Interferometer, \cite{cit.ISI}) at $11.15~\mathrm{\mu m}$ 
as published by \cite{cit.danchi1994}. 

A journal of all used data points is given in Table~\ref{tab.jrnl}. 
	\begin{table*}
	\centering
	\caption{\label{tab.jrnl}
	Journal of interferometric data. 
	K band (FLUOR) data for $\alpha$~Orionis have been presented in \cite{cit.perrin2002aori}, for SW~Virginis in \cite{cit.perrin2003a}, and for R~Leonis in \cite{cit.perrin1999}. 
	All L band (TISIS) data were taken from \cite{cit.chagnon2002}. 
	The ISI data points for $\alpha$~Orionis and R~Leonis were extracted from \cite{cit.danchi1994} using the DEXTER online tool at \cite{ads}. No exact position angle was given along with the ISI data, therefore we indicated the basic geographic orientation of the baselines used. 
	Where known for the objects, the phase $\Phi$ of the visual variability was indicated. 
	}
	\begin{tabular}{cccc|cccc}
	\hline\hline
  Object  &  Spatial Frq.  &  Visibility  &  Pos. Angle  &  Object  &  Spatial Frq.  &  Visibility  &  Pos. Angle  \\
    &  ($10^{5}~\mathrm{rad}^{-1}$)  &    &   (deg)   &    &  ($10^{5}~\mathrm{rad}^{-1}$)  &    &   (deg)   \\
	\hline

$\alpha$~Ori  &  \multicolumn{3}{l|}{FLUOR}  &          R~Leo  &  \multicolumn{3}{l}{FLUOR, Apr. 1996, $\Phi = 0.24$}  \\
  &  46.41  &  0.2232  $\pm$  0.0018  &  85.91  &	   &  65.76  &  0.2723  $\pm$  0.0040  &  86.9  \\ 
  &  46.82  &  0.2045  $\pm$  0.0020  &  80.47  &	   &  65.76  &  0.2718  $\pm$  0.0043  &  92.5  \\
  &  48.14  &  0.1847  $\pm$  0.0024  &  71.96  &	   &  66.38  &  0.2723  $\pm$  0.0041  &  80.3  \\
  &  50.39  &  0.1253  $\pm$  0.0016  &  63.01  &	   &  69.82  &  0.2373  $\pm$  0.0033  &  65.4  \\
  &  88.30  &  0.0949  $\pm$  0.0047  &  69.89  &	   &  71.57  &  0.2127  $\pm$  0.0035  &  60.7  \\
  &  90.92  &  0.0843  $\pm$  0.0024  &   --    &	   &  \multicolumn{3}{l}{FLUOR, Mar. 1997, $\Phi = 0.28$}  \\
  & 133.99  &  0.0510  $\pm$  0.0010  &  79.06  &	   &  47.94  &  0.4556  $\pm$  0.0041  &  86.0  \\
  &  \multicolumn{3}{l|}{TISIS, March 2000}  &              &  50.08  &  0.4248  $\pm$  0.0036  &  68.5  \\
  &  35.99  &  0.517  $\pm$  0.015  &   88.85  &           &  66.89  &  0.1934  $\pm$  0.0036  &  77.0  \\
  &  35.99  &  0.528  $\pm$  0.014  &   89.49  &           &  68.27  &  0.1751  $\pm$  0.0029  &  70.9  \\
  &  36.34  &  0.521  $\pm$  0.011  &   81.08  &           &  70.69  &  0.1656  $\pm$  0.0050  &  63.2  \\
  &  53.32  &  0.085  $\pm$  0.001  &  108.69  &           &  70.96  &  0.1761  $\pm$  0.0071  &  62.5  \\
  &  53.71  &  0.081  $\pm$  0.001  &  110.40  &           &  91.07  &  0.1255  $\pm$  0.0096  &  71.2  \\
  &  \multicolumn{3}{l|}{ISI, Oct. 1988 - Oct. 1989}  &   &  97.21  &  0.1188  $\pm$  0.0060  &  58.5  \\
  &  1.64  &  0.54  $\pm$  0.04  &  90  &                  & 150.62  &  0.0770  $\pm$  0.0032  &  63.1  \\
  &  1.82  &  0.57  $\pm$  0.02  &  90  &                  &  \multicolumn{3}{l}{TISIS, Mar. 2000, $\Phi = 0.81$}  \\
  &  2.24  &  0.61  $\pm$  0.01  &  90  &                  &  39.29  &  0.515  $\pm$  0.011  &  168.48  \\
  &  2.36  &  0.65  $\pm$  0.01  &  90  &                  &  39.46  &  0.528  $\pm$  0.011  &   67.44  \\
  &  2.51  &  0.62  $\pm$  0.02  &  90  &                  &  40.35  &  0.504  $\pm$  0.013  &   62.78  \\
  &  2.75  &  0.59  $\pm$  0.01  &  90  &                  &  52.76  &  0.249  $\pm$  0.002  &   86.66  \\
  &  2.96  &  0.59  $\pm$  0.01  &  90  &                  &  52.91  &  0.253  $\pm$  0.003  &   96.07  \\
  &  3.26  &  0.56  $\pm$  0.02  &  90  &                  &  53.69  &  0.258  $\pm$  0.003  &  103.86  \\
  &  3.53  &  0.60  $\pm$  0.01  &  90  &                  &  53.88  &  0.245  $\pm$  0.002  &  105.20  \\
  &  \multicolumn{3}{l|}{ISI, Oct. - Nov. 1992}  &        &  53.94  &  0.251  $\pm$  0.003  &  105.55  \\
  &  6.91  &  0.58  $\pm$  0.01  &  123  &                 &  54.04  &  0.246  $\pm$  0.002  &  106.15  \\
  &  7.39  &  0.59  $\pm$  0.01  &  123  &                 &  54.10  &  0.248  $\pm$  0.002  &  106.58  \\
  &  7.57  &  0.56  $\pm$  0.02  &  123  &		   &  54.12  &  0.227  $\pm$  0.002  &  106.74  \\
  &  7.96  &  0.54  $\pm$  0.02  &  123  &		   &  54.85  &  0.229  $\pm$  0.003  &  110.85  \\
  &  8.67  &  0.56  $\pm$  0.02  &  123  &		   &  \multicolumn{3}{l}{ISI, Oct. 1988, $\Phi \approx 0.45$}  \\
  &  \multicolumn{3}{l|}{ISI, Nov. 1989 - Oct. 1992}  &   &  2.44  &  0.86  $\pm$  0.05  &  90  \\
  &   8.97  &  0.61  $\pm$  0.02  &  113?  &		   &  2.54  &  0.83  $\pm$  0.04  &  90  \\
  &   9.18  &  0.58  $\pm$  0.01  &  113  &		   &  \multicolumn{3}{l}{ISI, Oct. 1990, $\Phi  = 0.81$}  \\
  &   9.30  &  0.60  $\pm$  0.02  &  113  &		   &  7.30  &  0.55  $\pm$  0.06  &  113  \\
  &   9.78  &  0.56  $\pm$  0.03  &  113  &		   &  7.51  &  0.63  $\pm$  0.03  &  113  \\
  &   9.96  &  0.55  $\pm$  0.02  &  113  &		   &  8.02  &  0.53  $\pm$  0.04  &  113  \\
  &  10.22  &  0.54  $\pm$  0.02  &  113  &		   &  8.23  &  0.65  $\pm$  0.05  &  113  \\
  &  10.55  &  0.56  $\pm$  0.02  &  113  &		   &  8.62  &  0.59  $\pm$  0.03  &  113  \\
  &  10.70  &  0.59  $\pm$  0.02  &  113  &		   &  9.10  &  0.54  $\pm$  0.03  &  113  \\
  &  10.91  &  0.54  $\pm$  0.01  &  113  &		   &  9.30  &  0.65  $\pm$  0.06  &  113  \\
  &  10.97  &  0.58  $\pm$  0.02  &  113  &		   &  9.61  &  0.53  $\pm$  0.03  &  113  \\
  &  11.30  &  0.54  $\pm$  0.01  &  113  &		   &  9.81  &  0.59  $\pm$  0.03  &  113  \\
  &  11.45  &  0.48  $\pm$  0.02  &  113  &		   & 10.02  &  0.60  $\pm$  0.02  &  113  \\
  &  11.57  &  0.55  $\pm$  0.02  &  113  &		   & 10.20  &  0.56  $\pm$  0.03  &  113  \\
  &  11.69  &  0.50  $\pm$  0.02  &  113  &		   & 10.32  &  0.57 \mbox{\rule[0ex]{6ex}{0ex}} &  113  \\
  &  11.81  &  0.52  $\pm$  0.01  &  113  \\
\cline{1-4}
SW~Vir  &  \multicolumn{3}{l|}{FLUOR, May 2000, $\Phi = 0.75$}  \\
  &  64.29  &  0.697  $\pm$  0.010  &  61.31  \\
  &  81.89  &  0.522  $\pm$  0.014  &  93.93  \\
  &  81.93  &  0.527  $\pm$  0.014  &  94.23  \\
  &  \multicolumn{3}{l|}{TISIS, Mar. 2000, $\Phi = 0.35$}  \\
  &  32.63  &  0.830  $\pm$  0.012  &  76.01  \\
  &  32.73  &  0.848  $\pm$  0.012  &  75.39  \\
  &  46.80  &  0.706  $\pm$  0.008  &  98.57  \\
\hline
	\end{tabular}
	\end{table*}

\subsection{Uniform disk diameters}\label{sect.ud-diam}
The visibilities $V_i$, and their standard deviations $\sigma_i$, 
in the K and L bands were fitted 
by a uniform disk model by minimising the quantity: 
	\begin{equation}\label{equ.ud-chisqu}
\chi^{2} = \sum_{i=1}^{n} 
	\frac{\left(V_{i}-M({{\phi_\mathrm{UD}},S_{i}})\right)^{2}}{\sigma_{i}^{2}} 
	\end{equation}
where $n$ is the number of available measurements in the respective band. 
$M$ is the uniform disk model:
	\begin{equation}\label{equ.ud-mod}
M({{\phi_\mathrm{UD}}, S_{i}}) = 
	\left|\frac{2J_{1}(\pi\phi_\mathrm{UD} S_{i})}{\pi\phi_\mathrm{UD}S_{i}}\right| ,
	\end{equation}
with $S_{i}$ the spatial frequency and $\phi_\mathrm{UD}$ the uniform disk
diameter. 
$J_1$ is the Bessel function of first order. 

The best fit uniform disk radii, and standard deviations, 
are listed in Table~\ref{resume}.
	\begin{table}
	\centering
	\caption{\label{resume}
	Uniform disk radii $\phi$ in the L and K bands.}
	\begin{tabular}{cccc}
	\hline\hline
	Star & $\phi_\mathrm{L}$ & $\phi_\mathrm{K}$ & 
		$\phi_\mathrm{L}/\phi_\mathrm{K}$ \\
	     & (mas) & (mas) &   \\
	\hline
$\alpha$~Ori & $21.37\pm 0.01$ & $21.63\pm 0.02$ & $0.99\pm 0.001$ \\
SW~Vir	     & $11.44\pm 0.17$ & {\rule[0ex]{1ex}{0ex}}$8.64\pm 0.16$ & $1.32\pm 0.030$ \\
R~Leo        & $17.97\pm 0.02$ & $14.72\pm 0.03$ & $1.22\pm 0.002$ \\
	\hline
	\end{tabular}
	\end{table}
The last column is the ratio of the L band and K band diameters. 
It is very close to 1 in the case of Betelgeuse, 
but the difference is 20--30\% for the other two stars. 

In the next sections, we investigate the possibility that 
dust envelopes around the star 
may be the cause for the diameter variations from K to L.

\section{Modelling of the dust envelope}\label{sect.model}
The %
models used in this paper are based on a radiative transfer code 
maintained at the Observatoire de la C\^{o}te d'Azur in Nice, France. 
A description of the code is presented by \cite{cit.niccolini2003}. 
We discuss hereafter the main hypotheses and parameters of this model.

\subsection{Description of the radiative transfer code}\label{sect.modelcode}
The code is based on the modelling 
of stellar radiation and circumstellar dust envelope interaction 
by a Monte Carlo method, 
assuming thermal equilibrium. 
The star is assumed to be a blackbody 
whose emitted energy is split in a number of packets, or ``photons''. 
These photons can either pass the envelope directly, 
or they are scattered once or more in the envelope, 
or the photons are absorbed and then thermally re-emitted by the dust grains. 
A grid over the dust envelope is defined with radial zones and angular sections, 
with the resolution accounting for expected temperature gradients. 
From the resulting temperature profile, 
the code derives the spectral energy distribution (SED) 
for the three individual photon flux contributions mentioned above. 
Initially, the stellar radius is left as scaling factor. 
Assigning a specific stellar radius, 
the total SED can be compared to photometric data 
in order to constrain the model.
Absorption, emission, and scattering by gaseous components
are not taken into account. 

Furthermore, the spatial flux distribution at a given wavelength can be derived, 
i.e., a model image of the object at this wavelength can be created. 
The Fourier transform of the model image can be compared to visibilities 
measured at that wavelength which yields a further constraint to the model. 
For the purpose of producing the image, the code allows a zooming factor 
while the image dimensions are always fixed at $300\times 300$ pixels.

\subsection{Parameters of the envelope}\label{sect.modelparams}
The code offers several parameters that can be chosen. 
The size distribution $s(a)$ of spherical dust particles 
having the radius $a$ was set as $s(a) \sim a^{-3.5}$, 
according to the interstellar medium \citep{cit.mathis1977} %
and independent of the distance from the star. 
The radius $a$ was contained in the range $0.005 - 0.25~\mathrm{\mu m}$. 

The particle density distribution $n(r)$ of the dust grains 
that was used in this study 
includes acceleration effects on dust by radiation pressure and gas drag 
(see Appendix~\ref{annexeA}). 
The resulting density distribution is much more peaked 
at the inner edge of the envelope 
than is a $r^{-2}$ distribution. 

For our study, the dust particle material was constrained to 
astronomical silicates. 
Their complex dielectric function (respectively complex refractive index) 
is given by \cite{cit.drainelee1984}. 
It was sampled at 37 wavelengths between $0.1~\mathrm{\mu m}$ and $300~\mathrm{\mu m}$, 
with the sample points considering local extrema. 

The previously mentioned parameters 
were chosen in consistency with the relevant subset of studies reported by \cite{cit.danchi1994}. 
They were set and fixed in this study. 
To fit the model to the data, the following input parameters were adjusted. 
The stellar radius $R_\star$ and  
the stellar effective temperature $T_\star$ 
govern the flux radiated by the star. 
Spherical symmetry was assumed for a dust shell around the star 
with an inner radius $r_\mathrm{i}$ and an outer radius $r_\mathrm{o}$, 
in units of $R_\star$. 
Furthermore, the envelope is characterised by its optical depth, $\tau_\lambda$, 
which is defined for overall extinction (absorption and scattering). 
It can be chosen at a given wavelength, $\lambda$, expressed in $\mu$m. 

Since the code, as used for this study, 
does not take into account possible changes of the dust properties 
when heated beyond the dust condensation temperature, 
the temperature map has to be checked accordingly for each set of parameters. 
As condensation temperature for silicates, 
we adopted $T_\mathrm{cond}\approx 1000~\mathrm{K}$ \citep{cit.gail1998}.

\section{Comparison between model and observations}
\label{sect.comp}
We now study the case of three particular stars. 
The question is whether a simple spherical dust shell model 
can account for the large diameter differences observed in the K and L bands. 
The goal is to find the stellar parameters 
for which visibility points in the K and L bands and at $11~\mathrm{\mu m}$ 
are compatible and simultaneously in agreement with the photometric data.

\subsection{A supergiant: $\alpha$~Orionis}\label{sect.alfori}
Betelgeuse is a Semi-Regular (SRc) variable star of spectral type M1I. 
It is a late-type supergiant star with a 2335-day period (see Table~\ref{Starscaract}). 

The mass-loss rate $\dot{M}$ of $\alpha$~Orionis is moderate.
\cite{cit.knapp1998} estimate a mass-loss rate from CO emission measurements 
of $3.1\cdot 10^{-7} M_{\sun}~\mathrm{yr}^{-1}$ at HIPPARCOS distance 
and with an outflow velocity $v_\mathrm{o}=14.2~\mathrm{km}~\mathrm{s}^{-1}$. 
Throughout this study, we adopt terminal outflow velocities from their work. 
For a distance $d$ of $150~\mathrm{p }$ 
and an outflow velocity $v_\mathrm{o}=15~\mathrm{km}~\mathrm{s}^{-1}$, 
\cite{cit.danchi1994} yield a mass-loss rate of $17\cdot 10^{-7} M_{\sun}~\mathrm{yr}^{-1}$. 
With the HIPPARCOS parallax of $7.63~\mathrm{mas}$ yielding a distance of $131~\mathrm{pc}$, 
and $\dot{M}\propto v_\mathrm{o}\, d$, 
their mass-loss rate is updated to $14\cdot 10^{-7}~M_{\sun}~\mathrm{yr}^{-1}$, 
which is given in Table~\ref{masslosses}. 

	\begin{table}
	\centering
	\caption{\label{masslosses}
	Mass-loss rate comparison. 
	The values for $v_\mathrm{o}$ come from \cite{cit.knapp1998}. 
	Data from \citeauthor{cit.danchi1994} were re-evaluated at HIPPARCOS distance 
	(see Table~\ref{Starscaract}) 
	following $\dot{M}\propto v_\mathrm{o}\, d$.}
	\begin{tabular}{ccccc}
	\hline\hline
Star  &  $v_\mathrm{o}$
  &  \multicolumn{3}{c}{Mass-loss rate ($10^{-7}~M_{\sun}~\mathrm{yr}^{-1}$)}  \\
  &  ($\mathrm{km}~\mathrm{s}^{-1}$)  
  &  $\dot{M}_\mathrm{Knapp~et~al.}$  &  $\dot{M}_\mathrm{Danchi~et~al.}$
  &  $\dot{M}_\mathrm{Present~paper}$  \\
	\hline
$\alpha$~Ori  &  $14.2$  &  $3.1$  &  $14$  &  $17$\\
SW~Vir        &  $7.8$  &  $1.7$  &  --  &  $1.4$\\ 
R~Leo         &  $6.8$  &  $0.94$  &  $0.60$  &  $0.77$\\
	\hline
	\end{tabular}
	\end{table}

\paragraph{Observational data} 
We used low spectral resolution data from different catalogues 
available at the \cite{simbad} Astronomical Database 
to constrain the SED over a large range of wavelengths. 
The TD1 catalogue provides fluxes around $0.2~\mathrm{\mu m}$. 
UBV measurements are reported in the GEN and UBV catalogues. 
Fluxes in the UBVRIJHKL bands are taken from the JP11 Johnson catalogue.  
The 12, 25, 60, and $100~\mathrm{\mu m}$ fluxes are from the IRAS catalogue, 
and \cite{cit.danchi1994} provide flux curves from 8 to $22~\mathrm{\mu m}$. 

For Betelgeuse, having slight luminosity variations, 
we can easily compare visibility points from different epochs. 
This assumption will be fully justified by the fit we obtain. 

\paragraph{Parameters} 
\cite{cit.danchi1994} propose a spherical dust shell model to account for
ISI data in the thermal infrared. 
This shell is supposed to have a small extent ($r_\mathrm{i}=45.9\,R_\star$, $r_\mathrm{o}=48.2\,R_\star$) 
quite far away from the central star and 
to be optically very thin ($\tau_{11}=0.0065$). 
It seems to indicate an episodic mass-loss process, 
leading to an empty area between the star and its envelope. 
We used these parameters (see Table~\ref{tab.model}) 
to reproduce the shorter wavelength visibility points given by FLUOR and TISIS.
	\begin{table}
	\centering
	\caption{\label{tab.model} 
	Parameters adjusted for modelling (see Sect.~\ref{sect.modelparams}). 
	$T_\star$: stellar temperature; $R_\star$: stellar radius; 
	$r_\mathrm{i}, r_\mathrm{o}$: inner respectively outer radius of dust shell; 
	$\tau_{11}$: optical depth of shell at $11~\mathrm{\mu m}$. 
	In all cases, an accelerated dust density profile was used. }
	\begin{tabular}{cccccc}
	\hline\hline
	Object  &  $T_\star$  &  $R_\star$  &  $r_\mathrm{i}/R_\star$  &  $r_\mathrm{o}/R_\star$  &  $\tau_{11}$\\
	        &   (K)   &  (mas)  &  &  & \\
	\hline 
	$\alpha$~Ori 	& 3640 & 21.8  & 45.9 &   48.2 & 0.0065\\
	SW~Vir 		& 2800 &  8.65 & 15.0 & 1000.0 & 0.045 \\
	R~Leo  		& 2500 & 14.9  &  3.5 &  140.0 & 0.1   \\
	\hline
	\end{tabular} 
	\end{table}

The effective stellar temperature of $3640~\mathrm{K}$ used by \cite{cit.danchi1994} is
fully consistent with the evaluation by \cite{cit.perrin2002aori} giving $3640~\mathrm{K}$ and $3690~\mathrm{K}$ depending on models. 
The stellar radius used in that model is $21.8~\mathrm{mas}$, satisfactorily accounting for both the photometric data and interferometric data in the three bands. 
This radius is close to the uniform disk model radii in the K and L bands of $21.6~\mathrm{mas}$ and $21.4~\mathrm{mas}$, respectively. 

\paragraph{Results} 
	\begin{figure*}
	\centering
	\includegraphics[width=0.49\textwidth]{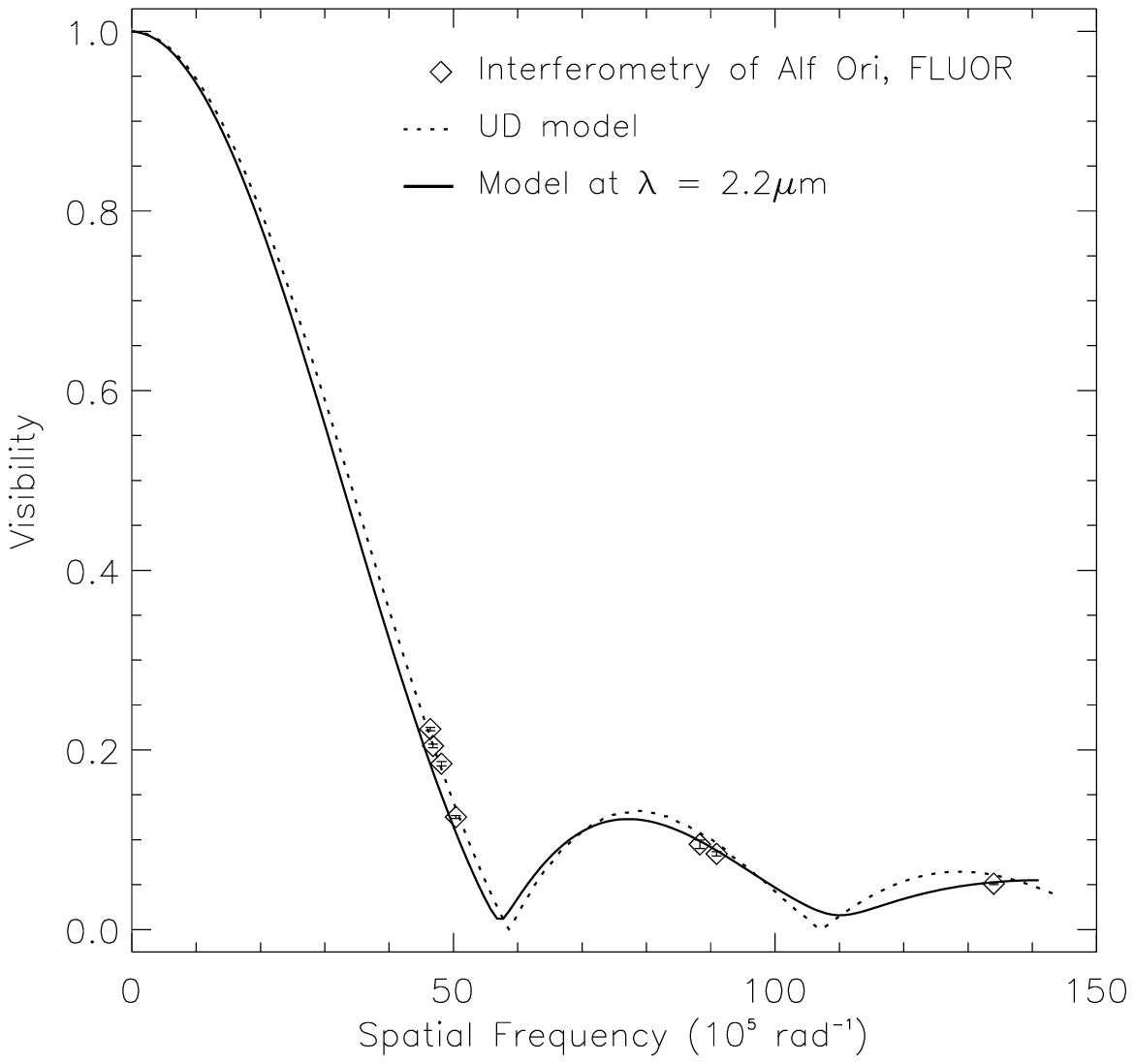} %
	\includegraphics[width=0.49\textwidth]{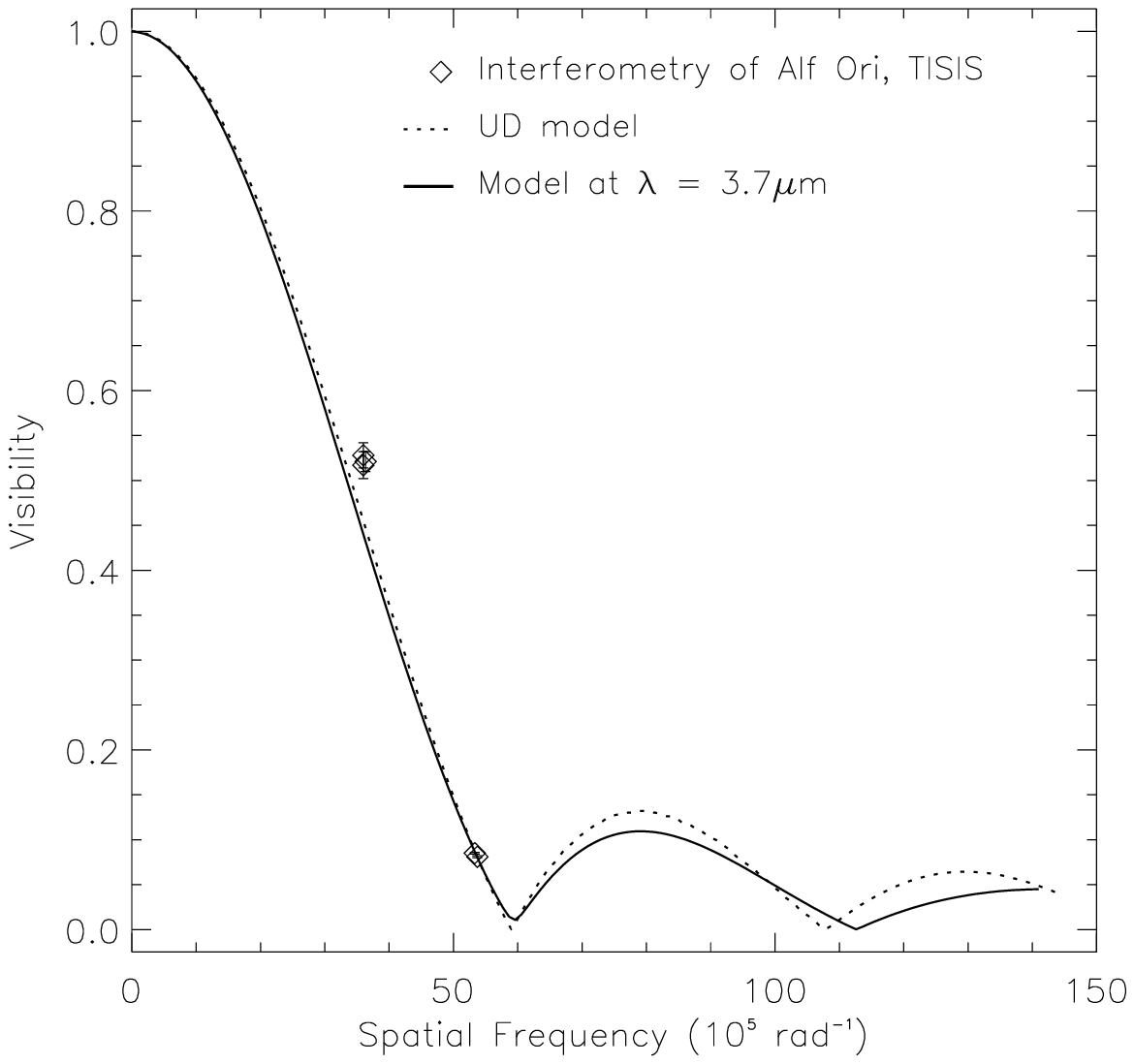}
	\\[3ex]
	\includegraphics[width=0.49\textwidth]{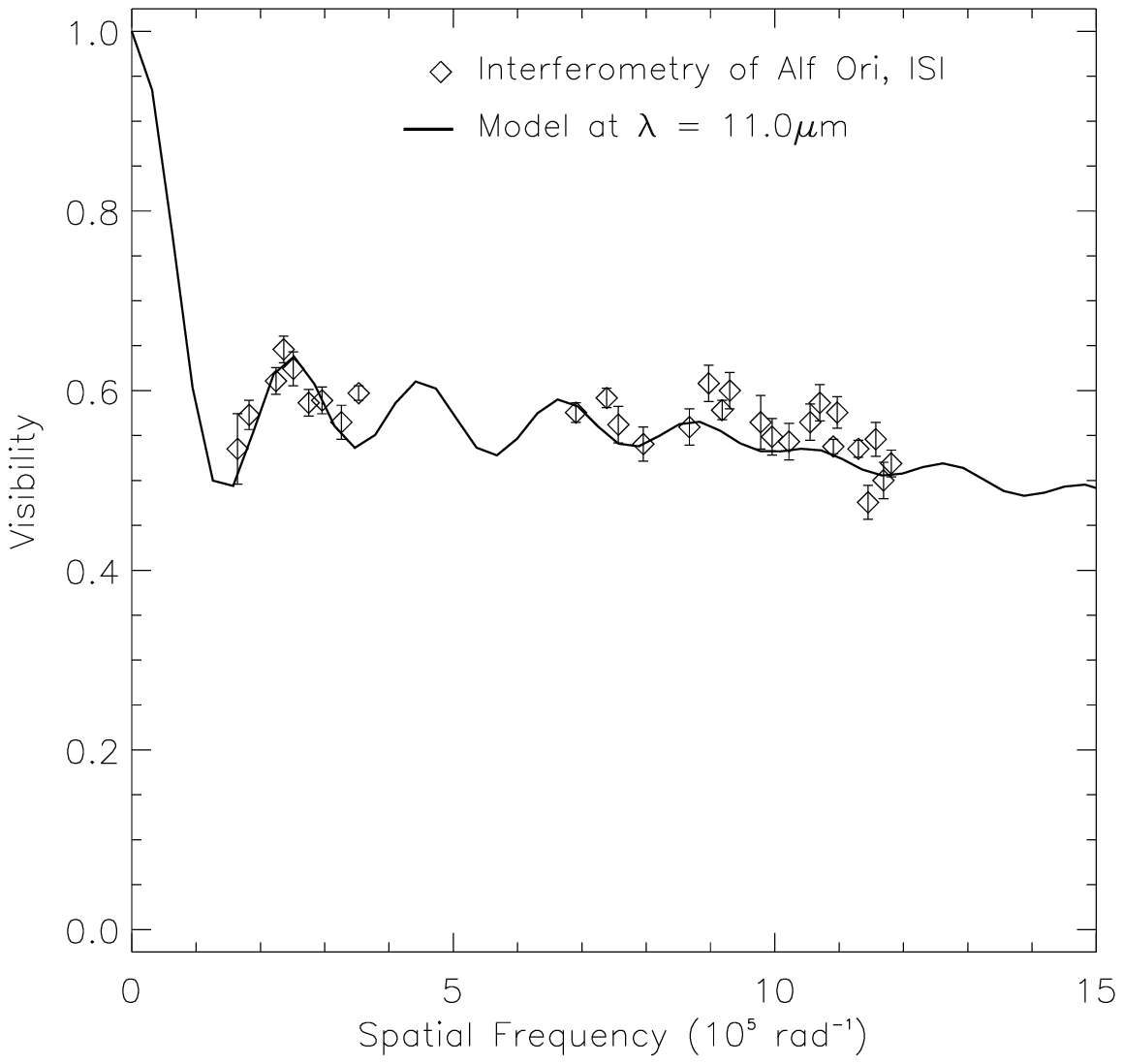}
	\includegraphics[width=0.49\textwidth]{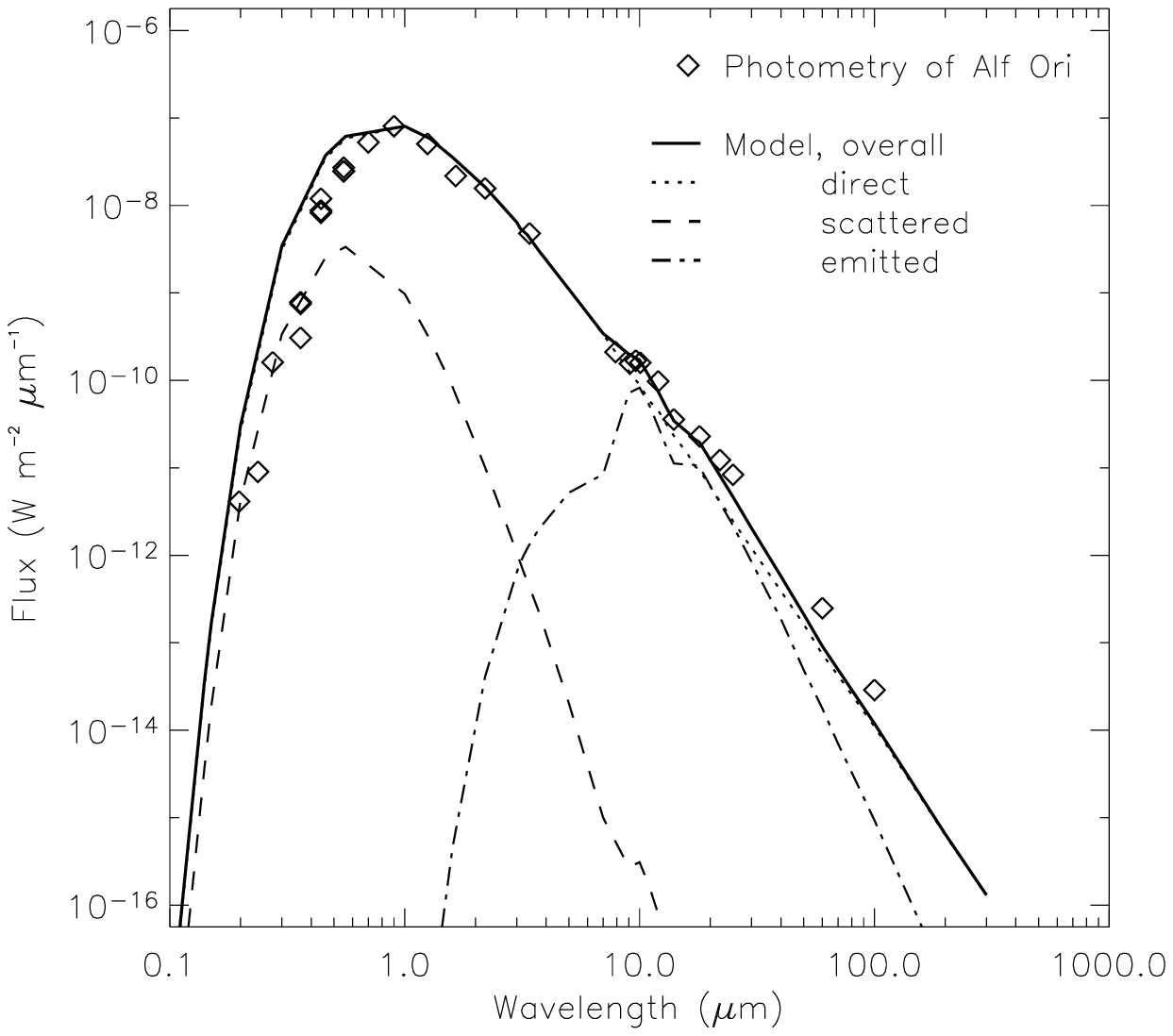}
\caption{\label{fig.output-alfori} 
Model output for $\alpha$~Ori. 
From top left to bottom left: modelled visibility curve with interferometric data for the bands K, L, and N; 
bottom right: modelled spectral energy distribution with photometric data. 
Model parameters (see Table~\ref{tab.model}): 
$T_\star=3640~\mathrm{K}$, $R_\star=21.8~\mathrm{mas}$, 
$r_\mathrm{i}=45.9\,R_\star$, $r_\mathrm{o}=48.2\,R_\star$, 
and $\tau_{11}=0.0065$. 
}
	\end{figure*}
The SED of the model is shown %
in Fig.~\ref{fig.output-alfori}.  
Apart from the total flux from the object, 
the plot also contains the contributions of the stellar flux attenuated by the circumstellar shell, 
the distribution of the flux scattered by the dust grains in the shell (maximum at short wavelength), 
and the thermal emission of the grains (maximum at longer wavelength). 
As the SED shows, stellar radiation is prominent over a large wavelength range. 
Flux emitted by dust becomes prominent only in a small band around $10~\mathrm{\mu m}$.

Visibility curves are also plotted in Fig.~\ref{fig.output-alfori}. 
The K and L band visibility curves from the dust shell model 
resemble each other 
and show good superposition 
with the respective visibility curves of the uniform disk model. 
Apparently, the same structure is seen in these two bands. 
The SED plot indicates that this should be the stellar photosphere 
since in these two bands 
the direct stellar flux is roughly $10^3$ times larger than the scattered part 
and about $10^5$ times larger than the dust emission. 
One can conclude that the dust envelope, 
having a very small optical depth, 
has little influence on the intensity distribution at these wavelengths. 
The dust shell is transparent in K and L 
and the star is seen through the dust at these wavelengths. 
As a consequence, 
the dust shell model derived from the ISI data is fully consistent
with the FLUOR and TISIS measurements.

As a further consistency check, yet with minor priority, 
we computed a mass-loss rate $\dot{M}$ from our model parameters 
by a method similar to \cite{cit.danchi1994} 
(see Appendix~\ref{annexeA} for details). 
With the set of parameters for $\alpha$~Ori, 
we obtain a mass-loss rate of $17\cdot 10^{-7}~M_{\sun}~\mathrm{yr}^{-1}$, 
consistent with \cite{cit.danchi1994} 
and roughly a factor of 5 higher than that of \cite{cit.knapp1998}. 

Although $\alpha$~Ori has the lowest optical depth of the studied objects, 
its mass-loss rate seems to be the highest. 
The parameters used here indicate an episodic event in the past. 
Yet, since this event the dust shell has expanded, 
with its enlarged inner radius entering the computation. 
For a further discussion of the mass-loss event see \cite{cit.danchi1994}.

We now study the case of stars with circumstellar envelopes at
higher optical depths.

\subsection{A semi-regular variable: SW~Virginis}\label{sect.swvir}
The pulsation period of this M7III semi-regular (SRb) variable star is 150 days.  
Its main characteristics are recapitulated in Table~\ref{Starscaract}. 
The mass-loss rate is estimated by \cite{cit.knapp1998} to $1.7\cdot10^{-7} M_{\sun}~\mathrm{yr}^{-1}$. 

\paragraph{Observational data}
Photometric data are retrieved from the IRC catalogue for the I and K bands, 
from the UBV catalogue in the respective filters, 
from the IRAS catalogue at 12, 60, $100~\mathrm{\mu m}$, 
and from \cite{cit.monnier1998} in the 8 to $13~\mathrm{\mu m}$ range. 
The M band flux at $5~\mathrm{\mu m}$ comes from \cite{cit.wannier1986}. 

To date, no 11-$\mathrm{\mu m}$ interferometric data is available for SW~Virginis. 
\cite{cit.benson1989} report observations by speckle interferometry in the N band 
at spatial frequencies below $2\cdot 10^{5}~\mathrm{rad^{-1}}$. 
Since they do not provide measured data directly, 
but rather give some model parameters, 
this has not been considered here. 
It is therefore not possible to assess the
consistency of the spatial model between all bands for this object. 

\paragraph{Parameters} 
For SW~Vir, \cite{cit.teff} give an effective temperature of $2921\pm 100~\mathrm{K}$, 
as derived from the bolometric flux and the limb-darkened diameter. 
Consistent with this value, 
the effective temperature was set to $2800~\mathrm{K}$.  
For the parameters of the dust shell 
we chose to adopt $r_\mathrm{i}=15.0\,R_\star$ and $r_\mathrm{o}=1000\,R_\star$, 
as proposed by \cite{cit.veen1995} on the basis of sub-mm observations. 
Applying an optical depth $\tau_{11}=0.045$ 
and assuming a stellar radius of $8.65~\mathrm{mas}$ 
then accounts for flux data at low spectral resolution. 

\cite{cit.danchi1994} make a distinction between two classes of objects.
The first class has a cold circumstellar envelope far from the star
(beyond 10 stellar radii), 
while objects of the second class are surrounded by a warmer dust envelope, 
near the star (at about 2 or 3 stellar radii). 
The mass-loss process is different for these two classes. 
For the second class, the mass-loss process 
is more continuous than for the first one. 
SW~Virginis has been classified in the past as a Mira star, 
hence with a somewhat regular mass-loss. 
Yet, its current classification as a semi-regular variable, 
possibly a precursor of a Mira star, 
makes its mass-loss more likely to be irregular, 
consistently with its irregular photometric variations. 
It is therefore very tempting to identify it as a member of the first class. 
The preferred set of parameters obtained for this star is 
summarised in Table~\ref{tab.model}. 
A large optical depth and a large shell extent are necessary 
to reproduce the photometric contribution of dust at large wavelengths.

\paragraph{Results} 
	\begin{figure*}
	\centering
	\includegraphics[width=0.49\textwidth]{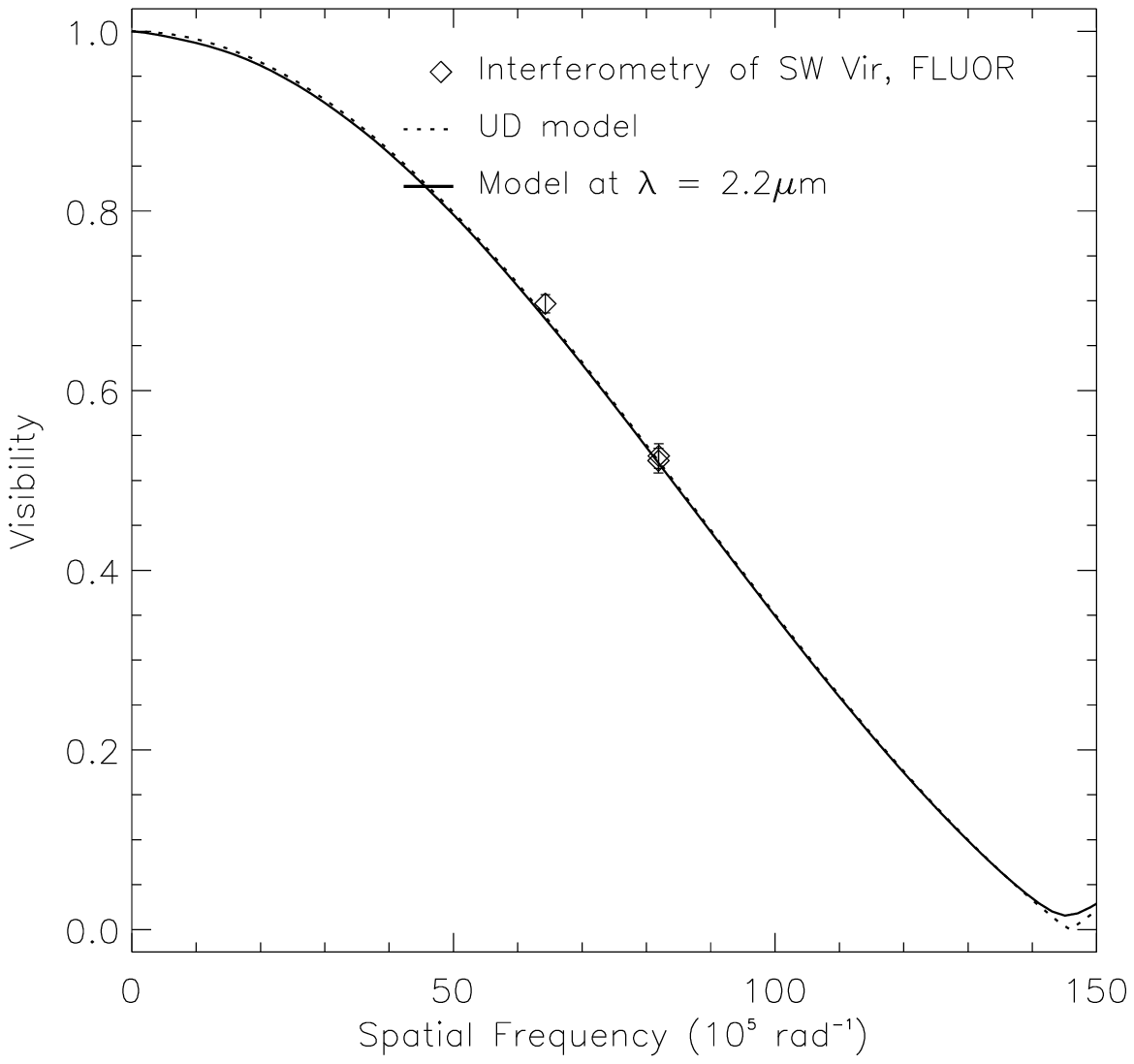} %
	\includegraphics[width=0.49\textwidth]{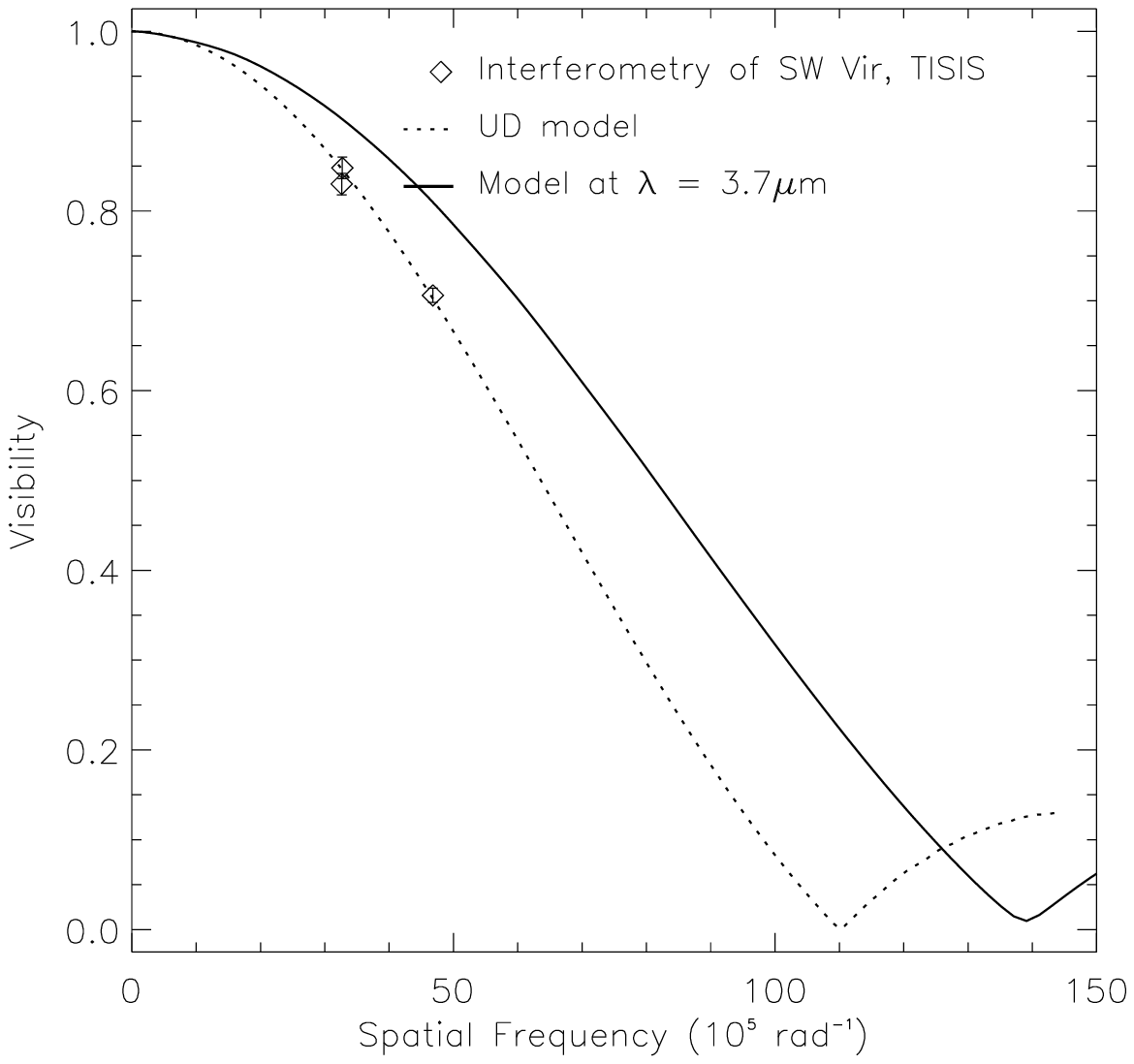}
	\\[3ex]
	\includegraphics[width=0.49\textwidth]{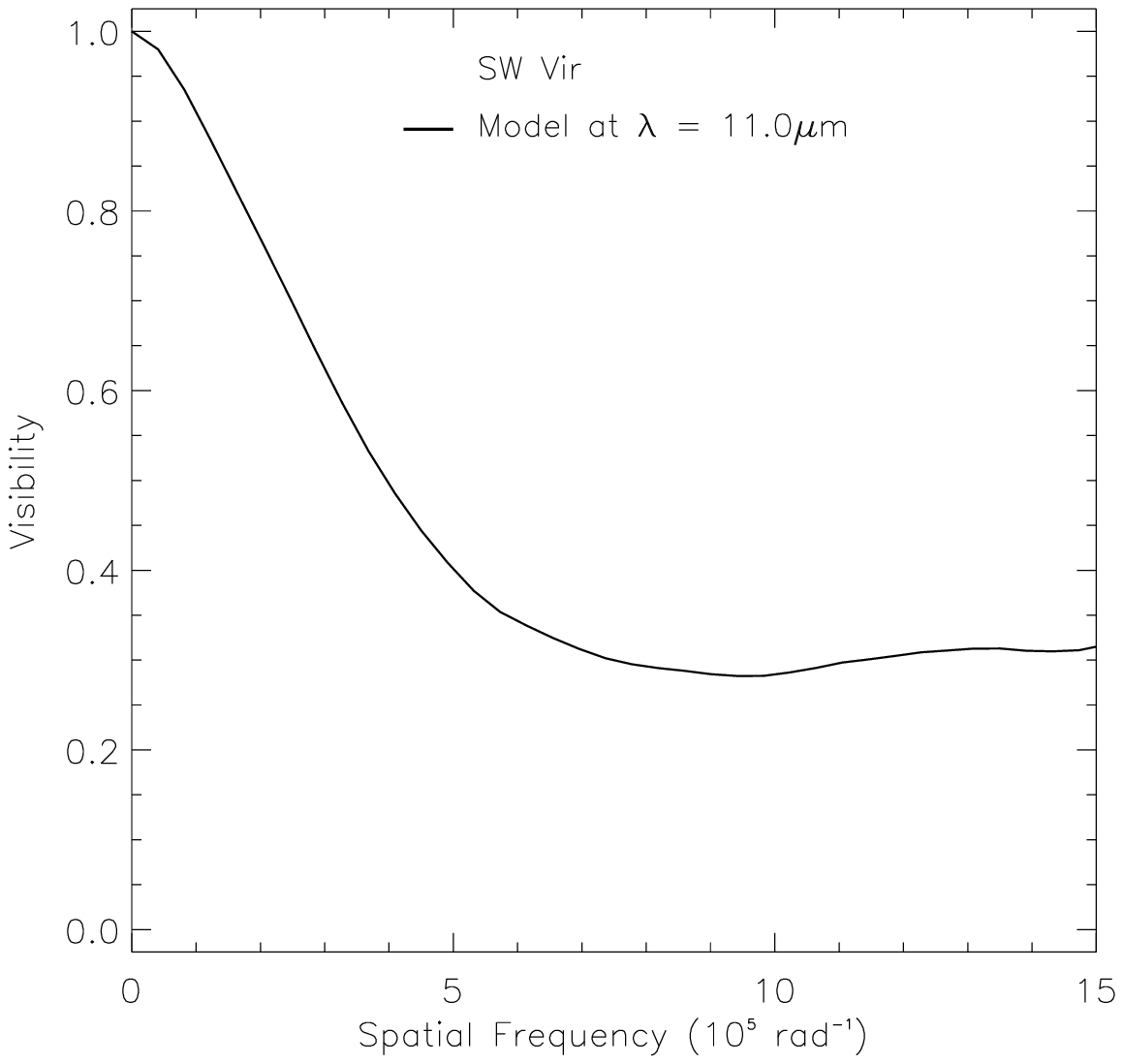}
	\includegraphics[width=0.49\textwidth]{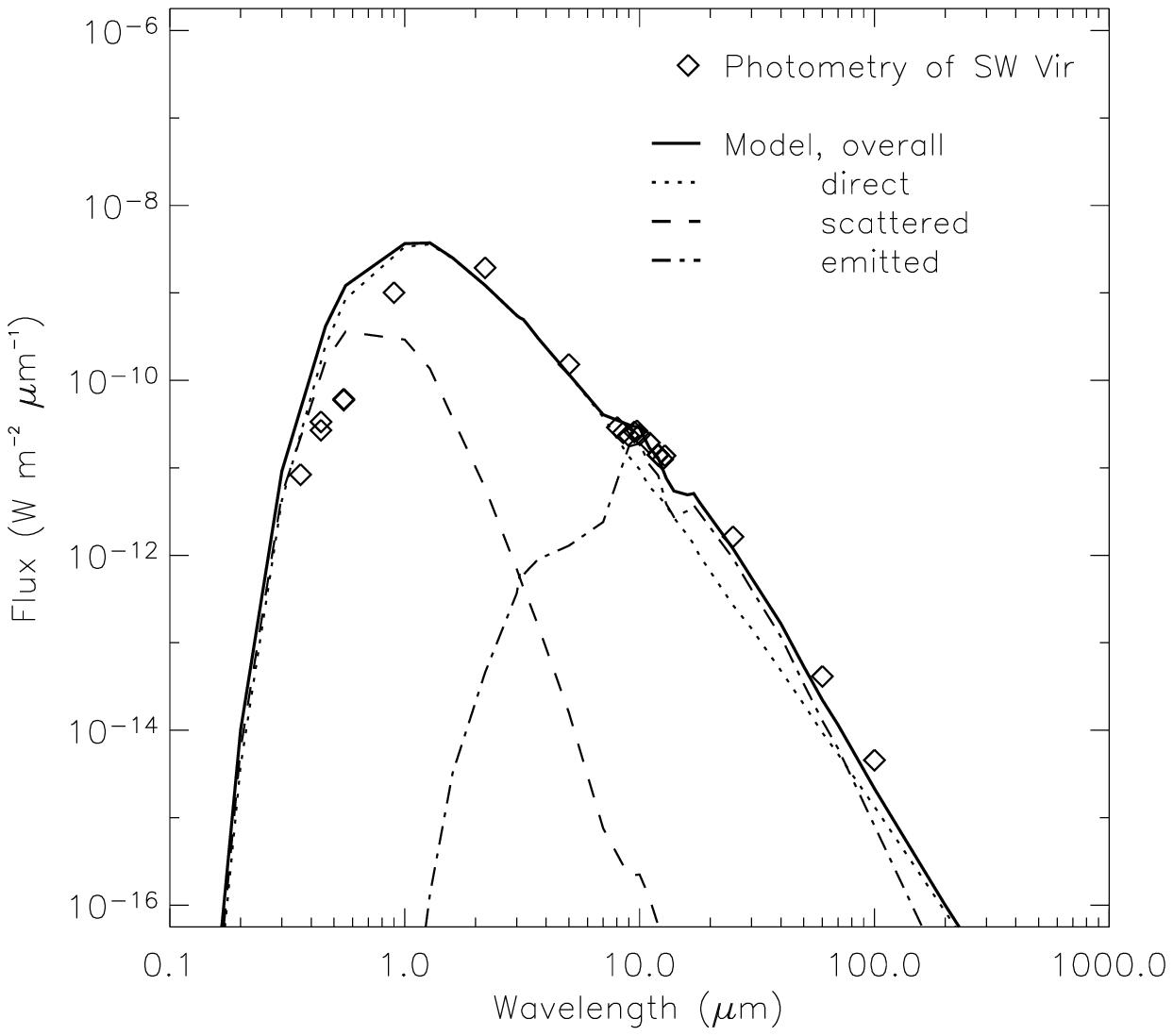}
\caption{\label{fig.output-swvir} 
Model output for SW~Vir. 
Top left and right: modelled visibility curve with interferometric data for the K and L bands; 
bottom left: expected visibility curve for the N band; 
bottom right: modelled spectral energy distribution with photometric data. 
Model parameters (see Table~\ref{tab.model}): 
$T_\star=2800~\mathrm{K}$, $R_\star=8.65~\mathrm{mas}$, 
$r_\mathrm{i}=15.0\,R_\star$,  $r_\mathrm{o}=1000\,R_\star$, 
and $\tau_{11}=0.045$. 
}
	\end{figure*}
As Fig.~\ref{fig.output-swvir} shows, 
the contribution of dust emission to the SED in the K and L bands of SW~Virginis 
is more important than in the case of $\alpha$~Ori, 
i.e., it is higher in relation to the direct flux. 
This can be attributed to the larger optical depth. 
SW~Virginis has some infrared excess that requires a dust shell. 
However, the dust cannot be too warm, otherwise it would produce a much larger
infrared excess than is compatible with the SED.

From this model, visibility curves were derived. 
Although comparatively few visibility measurements are available for this object, 
it is obvious that the single dust shell model 
is not simultaneously compatible with the K and L band data. 
Again, model visibility curves are similar in K and L 
which lets us assume that we see the same structure, i.e. the stellar photosphere. 
This is also reflected by the SED for which a similar reasoning applies 
like in the case of $\alpha$~Ori. 
Indeed, the dust being quite far away from the photosphere, 
the K band data are well reproduced by the visibility model. 
However, the dust is too cold and optically too thin 
to lead to a significant increase of
the diameter at larger wavelengths compatible with the L band data. 
A simple dust shell model, therefore, cannot account for both the
interferometric and the photometric data. 
Unfortunately, no ISI data are available for this star. 
They would have been very useful to help 
model the dust shell and place stronger constraints on it. 

The mass-loss rate we obtain from our parameters is $1.4\cdot 10^{-7}~M_{\sun}~\mathrm{yr}^{-1}$, 
which is roughly consistent with the one of \cite{cit.knapp1998}.

\subsection{A Mira-type variable: R~Leonis}\label{sect.rleo}
R~Leonis is a Mira-type long-period variable 
with a period of $\sim 310$ days 
and of late spectral type M8IIIe (see Table~\ref{Starscaract}). 
Its mass-loss rate is indicated by \cite{cit.knapp1998} 
as $0.94\cdot 10^{-7}~M_{\sun}~\mathrm{yr}^{-1}$, 
whereas the estimate by \cite{cit.danchi1994} can be updated to $0.60\cdot10^{-7} M_{\sun}~\mathrm{yr}^{-1}$. 

R~Leonis is known to have a very strong infrared excess 
and very steady pulsations which makes it very different from SW~Virginis. 
R~Leonis shows large photometric variations, especially at short wavelengths. 
The problem here is to synthesise a low spectral resolution SED, 
as well as visibility curves, 
for a star with large photometric variations.
Nevertheless, we decided to include all available data points 
(both photometric and interferometric) into the plots 
despite the object variability 
which will allow a larger range of acceptable model parameters.

\paragraph{Observational data} 
For R~Leonis, UBV measurements can be found in the GEN and UBV catalogues. 
The JP11 Johnson catalogue provides magnitudes in the BVRIJKL bands. 
\cite{cit.strecker1978} observed this object between 1.2 and $4.0~\mathrm{\mu m}$. 
The fluxes at 12, 20, 60 and $100~\mathrm{\mu m}$ are from IRAS, 
and fluxes between 8 and $20~\mathrm{\mu m}$ come from \cite{cit.danchi1994}.

\paragraph{Parameters}\label{par.rleo-param} 
The effective temperature that accounts for the photometric data is about $2500~\mathrm{K}$, 
a bit more than the temperature of $2000~\mathrm{K}$ assumed by \cite{cit.danchi1994} 
for their model containing only silicate dust around R~Leo. 
The stellar radius is clearly smaller ($14.9~\mathrm{mas}$ instead of $19.8~\mathrm{mas}$). 
However, we have estimated these two parameters with all data combined 
(interferometric data in the K and N bands and the spectro-photometric data), 
and they show good consistency 
(in as much as the star is neither a perfect blackbody nor a uniform disk) 
with those derived from the K band interferometric data alone by \cite{cit.perrin1999}, 
who give uniform disk radii of at most $\sim 15~\mathrm{mas}$ and photosphere
effective temperatures closer to $3000~\mathrm{K}$ rather than $2000~\mathrm{K}$. 
Our effective temperature is larger than that of \citeauthor{cit.danchi1994} 
which is not surprising 
since in their case only photometric data above $8~\mathrm{\mu m}$ was used, 
i.e., in a range of wavelengths very sensitive to cool dust emission. 
This also explains why their stellar radius is larger: 
for a given bolometric flux, 
the stellar diameter increases with decreasing temperature. 

As in \cite{cit.danchi1994}, we consider a large circumstellar envelope near
the central star. 
But instead of 2~$R_{\star}$ for the inner edge of the dust shell, 
we chose 3.5~$R_{\star}$ to keep the temperature of the grains close to $1000~\mathrm{K}$, 
the assumed temperature of grain condensation. 
The overall parameters are summarised in Table~\ref{tab.model}.

\paragraph{Results} 
	\begin{figure*}
	\centering
	\includegraphics[width=0.49\textwidth]{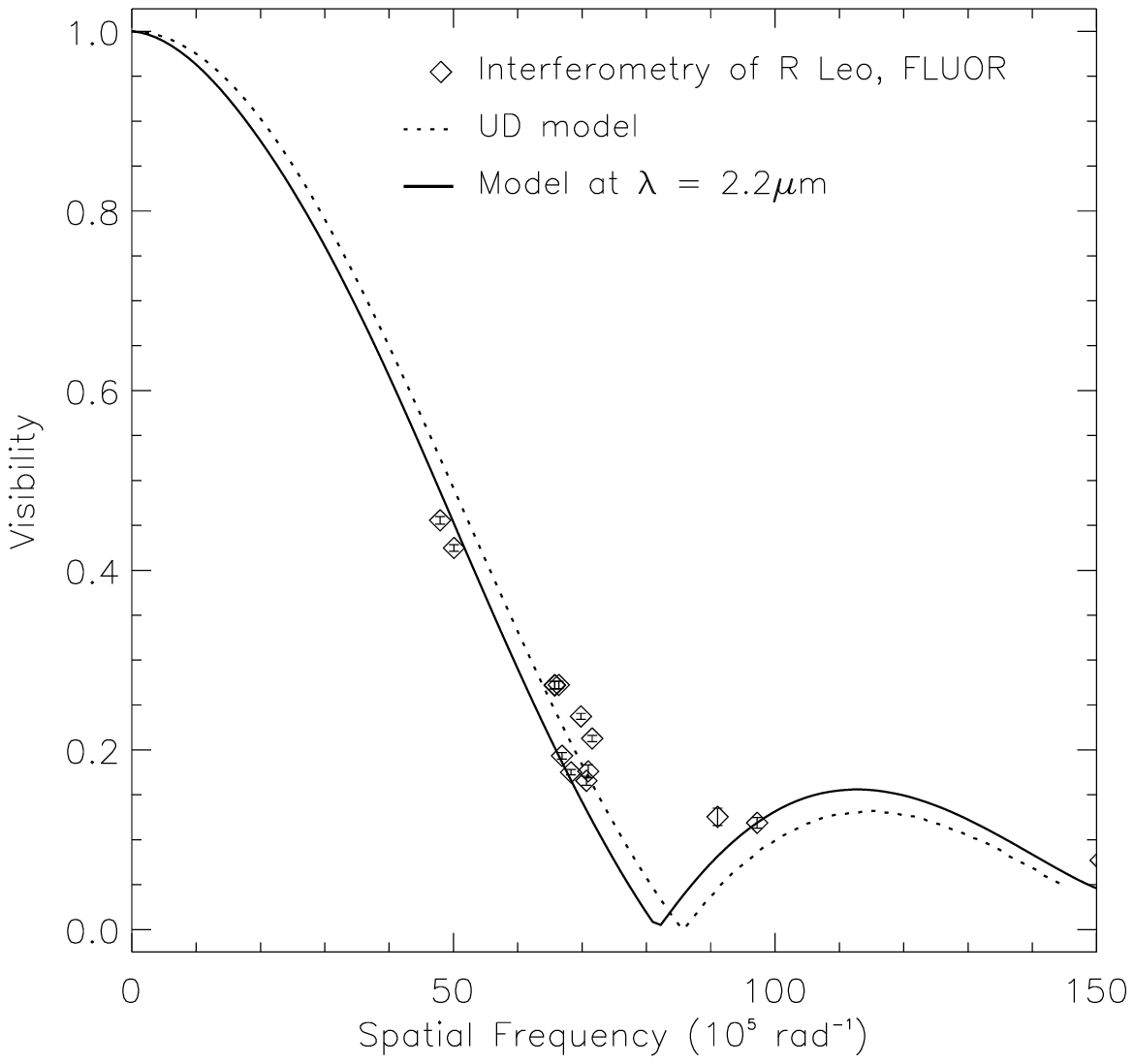} %
	\includegraphics[width=0.49\textwidth]{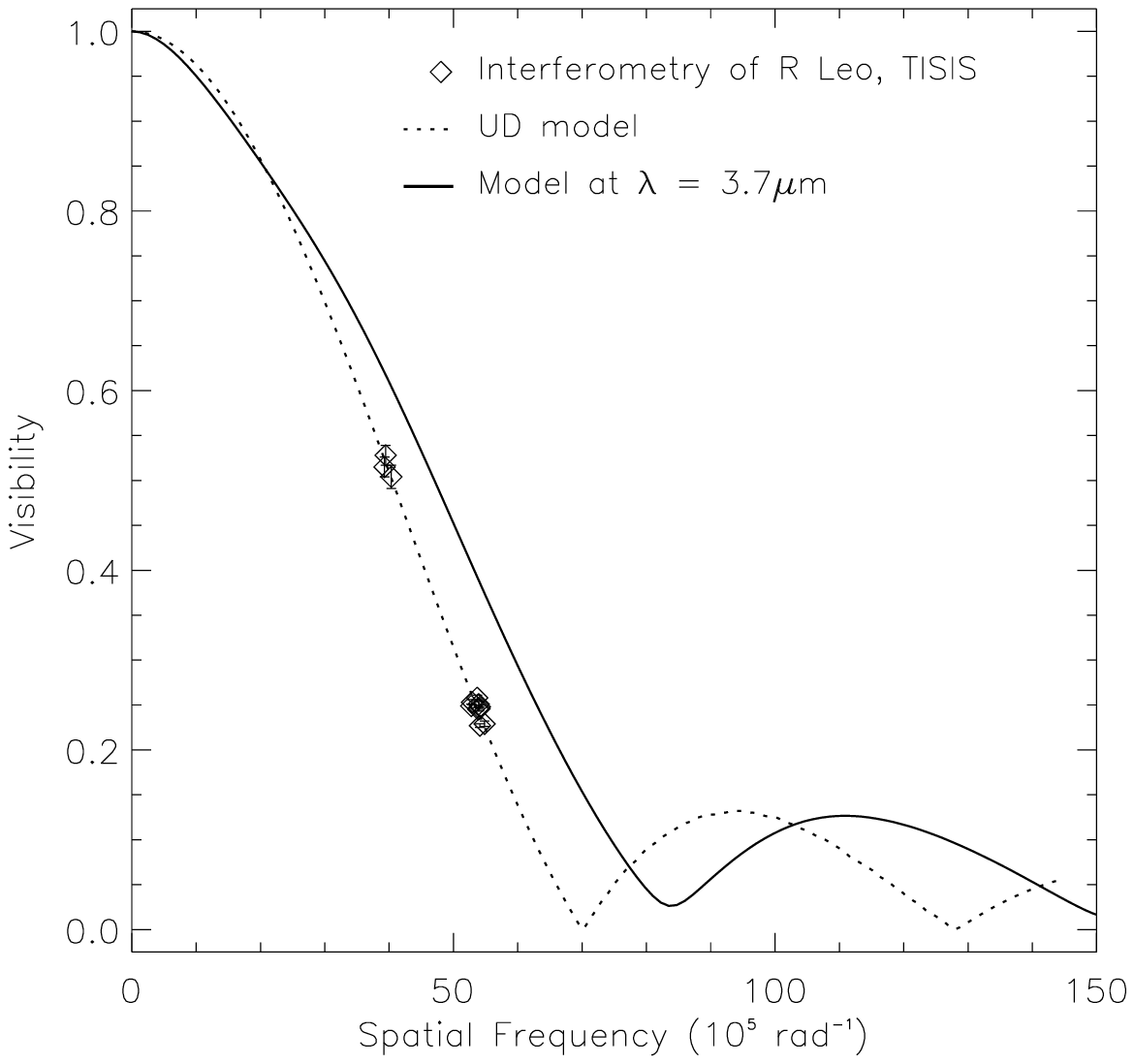}
	\\[3ex]
	\includegraphics[width=0.49\textwidth]{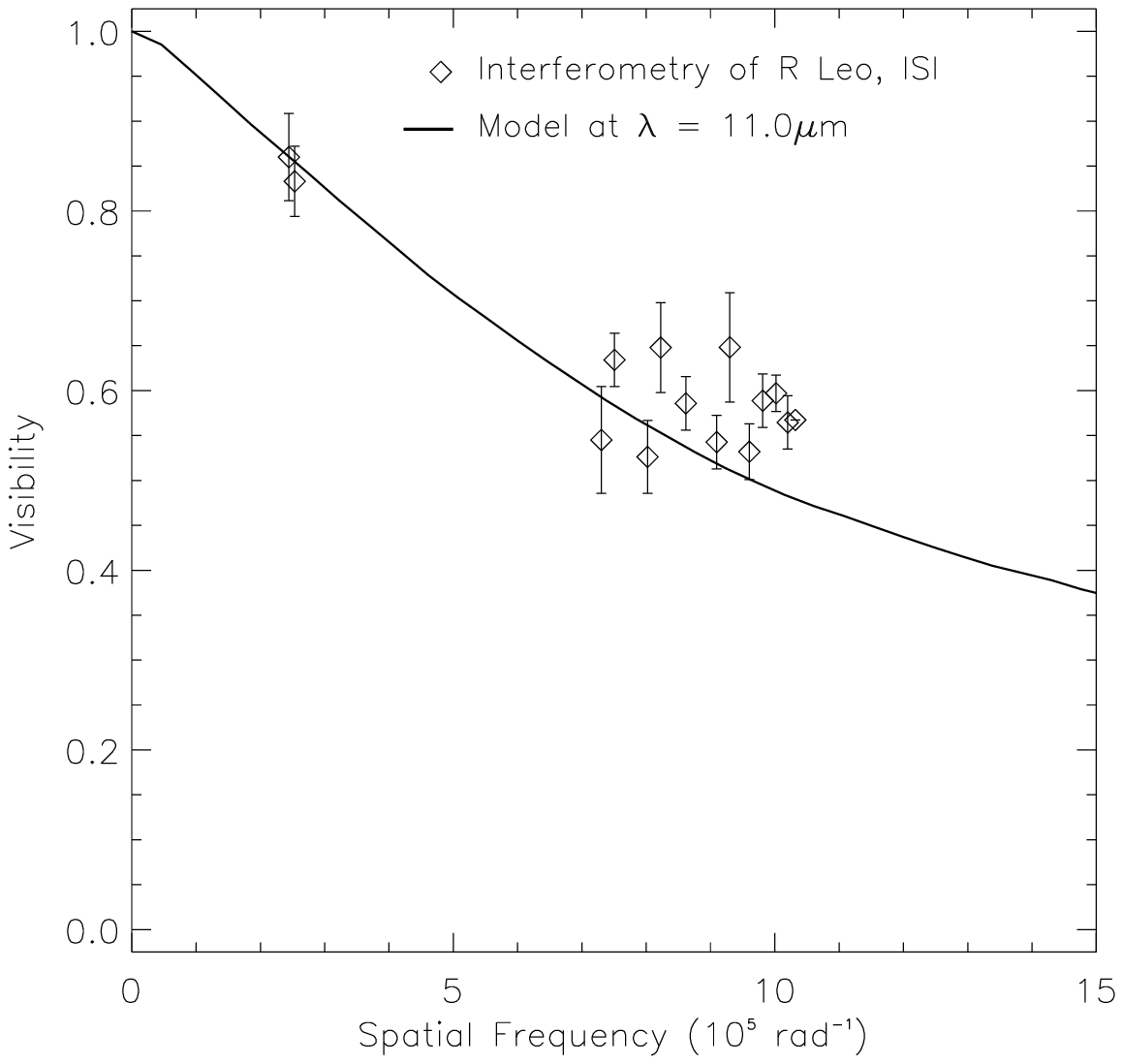}
	\includegraphics[width=0.49\textwidth]{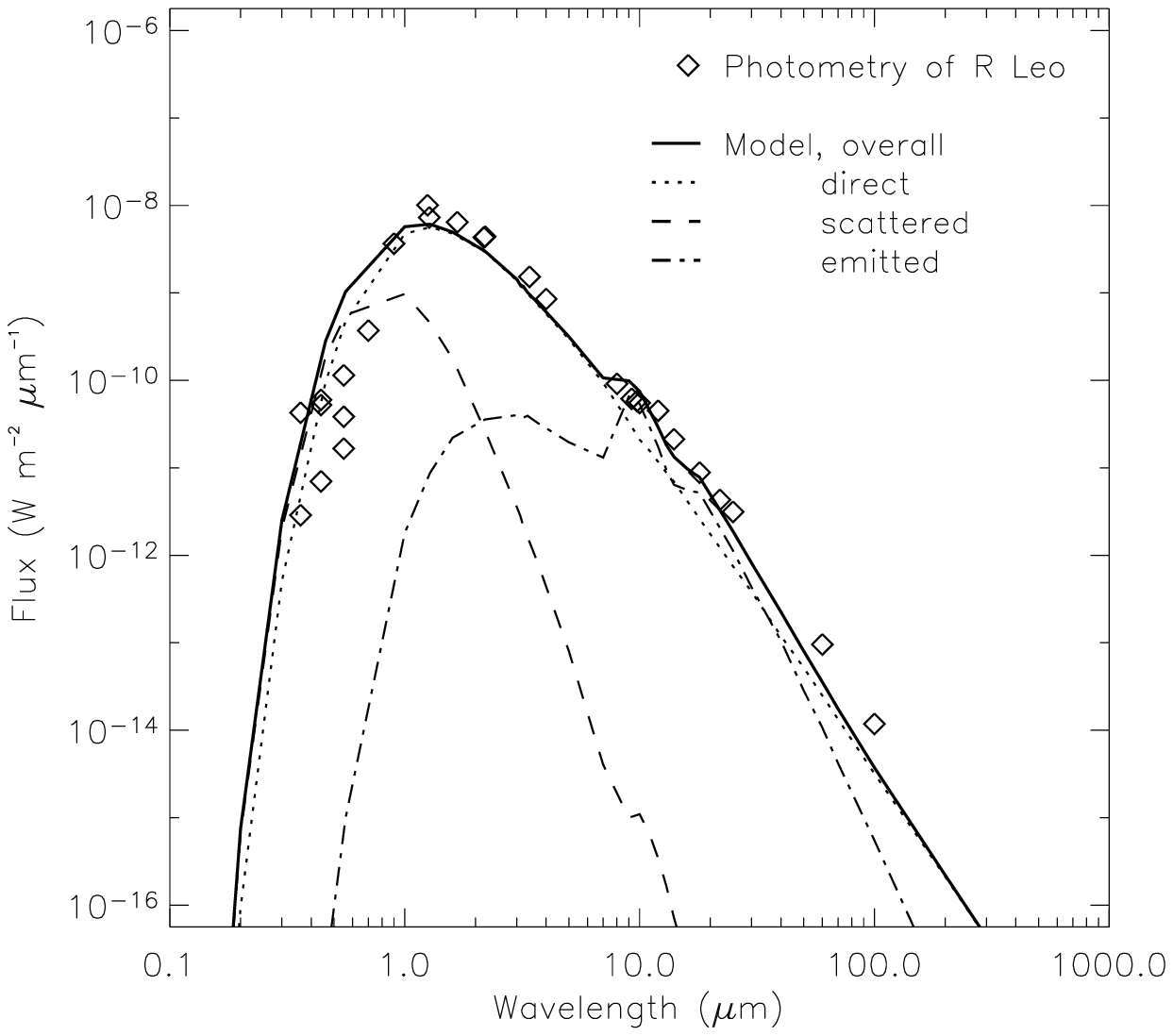}
\caption{\label{fig.output-rleo} 
Model output for R~Leo. 
From top left to bottom left: modelled visibility curve with interferometric data for the bands K, L, and N; 
bottom right: modelled spectral energy distribution with photometric data. 
Model parameters (see Table~\ref{tab.model}): 
$T_\star=2500~\mathrm{K}$,  $R_\star=14.9~\mathrm{mas}$, 
$r_\mathrm{i}=3.5\,R_\star$,  $r_\mathrm{o}=140\,R_\star$, 
and  $\tau_{11}=0.1$. 
}
	\end{figure*}
The fit of the photometric data described in the previous section is
displayed in Fig.~\ref{fig.output-rleo}. 
The contribution of the envelope dominates beyond $10~\mathrm{\mu m}$ up to about $30~\mathrm{\mu m}$. 
With a quite large extent, up to 140~$R_{\star}$, 
and an optical depth ${\tau}_{11}=0.1$
the dust shell accounts for IRAS and \cite{cit.danchi1994} observations in the
long wavelength range.  
As can be noticed in the graph, scattered flux and thermal dust emission 
are even more important than for SW~Vir. 
This is apparently due to the larger optical depth and the smaller inner radius of the envelope. 
These parameters lead to a mass-loss rate of $0.77\cdot 10^{-7}~M_{\sun}~\mathrm{yr}^{-1}$, 
which is of the same order of and intermediate between the values published 
by \cite{cit.danchi1994} and \cite{cit.knapp1998} (see Table~\ref{masslosses}). 

Visibility curves gained from this model are also shown 
in Fig.~\ref{fig.output-rleo} for the K, L, and N bands respectively. 
The stellar radius derived from our model is $14.9~\mathrm{mas}$, 
much smaller than the $19.8~\mathrm{mas}$ %
of \cite{cit.danchi1994}. 
Since our study is based on interferometric data 
spanning a wider range of wavelengths, 
the parameters we derive are different from that of \cite{cit.danchi1994} 
The ISI data are better suited to characterise the dust, 
yet the shorter-wavelength data bring more accurate stellar parameters, 
hence explaining the discrepancies between the two studies. 
The dust envelope structure may be considered as stationary 
with little sensitivity to stellar pulsation. 
FLUOR observations at phase 0.28 (March 1997) in K band 
give a uniform disk radius of $15.34~\mathrm{mas}$ \citep{cit.perrin1999}. 
In principle, the K band measurements are more sensitive 
to regions very close to the photosphere. 
It therefore seems that the radius of R~Leonis should be smaller than the 
evaluations by \cite{cit.danchi1994} and the difference between our estimate 
and that of \citeauthor{cit.danchi1994} cannot be attributed to the star pulsation. 
This difference must be attributed to the difference of effective temperatures. 

We now compare interferometric data in the K and L band and that of ISI. 
Our parameters are still consistent with the ISI visibility points 
at $11~\mathrm{\mu m}$ and account for FLUOR observations in the K band 
at low spatial frequencies. 
It is to be noticed that, 
since the ISI visibility data are at low spatial frequencies, 
they are not as sensitive to the different models 
as are the visibilities at shorter wavelengths. 
This explains why they are consistent with 
both the \cite{cit.danchi1994} and our model. 
On the other hand, in the L band our model overestimates 
the measured visibility values. 
As in the case of SW~Vir, 
it appears that the large structure seen in the L band 
is not well accounted for by a simple dust shell model, 
despite the even larger optical depth. 
Also the SED shows that contributions from the dust shell 
are at least a factor $10^2$ below the direct stellar flux. 
A larger radius could explain both $3.6~\mathrm{\mu m}$ and $11~\mathrm{\mu m}$ observations, 
but would disagree with K band measurements and the SED.

\section{Discussion}\label{sect.discuss}
\subsection{Obtained results}\label{sect.discuss-res}
From what we have seen so far, the best consistency between the 
multiple band data and the dust shell model is obtained for $\alpha$~Ori. 
The model described by \cite{cit.danchi1994} has a 
stellar radius close to the K and L radii estimated with a uniform disk model. 
Yet, in this case the influence of the envelope is not very important 
because of its small thickness and low opacity. 
Nevertheless, model parameters for ISI data from \citeauthor{cit.danchi1994} 
are consistent with what we see both for K and L bands. 

One may note, however, that 
\cite{cit.weiner2000} report further ISI data on $\alpha$~Ori 
which are not fully consistent with the model of \citeauthor{cit.danchi1994} 
based on observations at low spatial frequencies. 
\citeauthor{cit.weiner2000} took their measurements at higher spatial frequencies 
where the dust shell is fully resolved. 
Its contribution to the visibility is therefore negligible 
and the observed data are governed 
by radiation coming directly from the star or from its close environment. 
Like the L band data of SW~Vir and R~Leo (see below), 
the data of \citeauthor{cit.weiner2000} are overestimated by the dust shell model. 
These findings are further discussed in \cite{cit.perrin2002aori}. 

For AGB stars with later types like SW~Vir and R~Leo, 
for which we measure large diameter differences between $2.15~\mathrm{\mu m}$ and $3.77~\mathrm{\mu m}$ 
(the approximate effective wavelengths of the FLUOR and TISIS instruments), 
a simple dust shell model cannot account for these differences. 
However, the shape of near-infrared visibility curves for SW~Vir and R~Leo 
seem to show a trend. 
L band visibility points are systematically overestimated 
by a simple dust shell model 
meaning the source is larger than predicted by the model. 
This shows that the stellar environment mainly seen at this wavelength 
is much more complex than suggested by a model 
that takes into account only the star and its dust envelope. 

Indeed, the temperature of this area is close to the temperature 
at which grains form. 
Consequently, we must also take the presence of gas into account in modelling. 
Doing this will radically change the radiation transfer behaviour. 
Recently, spectroscopic detection of SiO, OH, and H$_2$O masers in late-type 
star envelopes have revealed the necessity of modelling the gas 
to explain large diameters seen in L band. 
ISO/SWS and other studies (\cite{cit.tsuji1997,cit.yamamura1999,cit.tsuji2000} brought up the existence 
of molecular layers of warm and cold H$_2$O, SiO, and CO$_2$ around AGB stars. 
Such layers in AGB stars certainly exist and have an impact 
on the stellar radiation which is observed by infrared interferometers. 
Strong molecular scattering by H$_2$O and CO were suggested 
to account for K band visibility measurements of R~Leo by FLUOR in 1997 \citep{cit.perrin1999} 
and seem to explain more easily the near-infrared measurements \citep{cit.mennesson2002, cit.perrin2003c}. 
It is very likely that dust needs to be coupled with gas 
to produce a more reliable model of these sources 
and that dust alone %
cannot be the only cause of the measurements taken in a wide range of wavelengths. 
The inclusion of a detailed photospheric model 
with line opacity due to gas 
would improve the agreement of the SED, particularly in the blue, 
and would improve the discrimination between stellar and shell components.

\subsection{Data used}\label{sect.discuss-data}
Certain aspects of our study can be further weighed. 
Simultaneous interferometric data would certainly help 
constrain models more accurately. 
Same epoch measurements are needed because of the large diameter
variations of AGB stars. 
At least for the position angles it can be noted, though, 
that they are fairly close in range, 
which leaves little possibility that the discrepancy we see 
may be explained by an elliptical shape (see below). 
Furthermore, 11.15-$\mathrm{\mu m}$ data would have been very useful 
to see whether our SW~Vir dust model parameters are correct 
since the envelope has more impact on the model at this wavelength. 
Such data might be provided in the near future 
by single-dish instruments on 10-m class telescopes or 
by upcoming mid-infrared interferometers like MIDI at the VLTI.

\subsection{Model code parameters}\label{sect.discuss-code}
A limitation of the model as used for this study is that 
it does not consider heterogeneous grain size distributions $s(a)$ 
depending upon the distance from the star 
or temporal effects (like grain formation, growth, or destruction). 
Also, different limits of the grain radius $a$ 
might have affected the fit of the SED in the UV and the far-infrared. 
Yet, varying the grain size range from $0.005 - 0.25~\mathrm{\mu m}$ 
to $0.005 - 2.5~\mathrm{\mu m}$ and $0.05 - 2.5~\mathrm{\mu m}$ for R~Leo as a study case, 
the resulting diagrams showed only small changes. 
It appeared that, in the modelled visibility curves for the K and L bands, 
the contribution of spatial frequencies below $20\cdot 10^5~\mathrm{rad}^{-1}$ 
were somewhat increased. 
Above all, neither the star nor the envelope are ideal blackbodies 
which adds to these deviations. 

Concerning the particle density distribution $n(r)$, 
\cite{cit.suh1999} questions the necessity 
of assuming a distribution which is peaked at the inner edge. 
Indeed, comparing results of calculations with two sets of parameters 
that differ only in $n(r)$ showed only minor changes in K and L bands. 
For $11.15~\mathrm{\mu m}$, changes were somewhat clearer, 
whereby for $\alpha$~Ori and R~Leo the accelerated version fitted ISI data better. 
Other issues are the possible existence of smaller structures in the envelope 
like clumps or aspherical dust distributions, as they were investigated, for example, by \cite{cit.lopez1997}. 
These have not been tested here and require more intense visibility modelling. 
This would be definitely required 
if sources showed systematic asymmetric effects, 
yet which are likely to 
produce ringing features on top of a visibility curve 
compared to the more basic effect of radiative transfer in spherical shells. 
Nevertheless, closure phase observations could help test for asymmetries. 

As for the dust particle material, 
slightly differing optical properties were proposed by several authors. 
See, for example, \cite{cit.suh1999} for a comparison of optical properties 
of silicate dust grains in the envelopes around AGB stars. 
We chose the ones of \cite{cit.drainelee1984} which are widely used. 
Using optical properties suggested by \cite{cit.suh1999} for comparison 
made it necessary to choose different dust shell parameters 
in order to achieve reasonable consistency with ISI data. 
Yet, L band data were still overestimated, 
similar to Fig.~\ref{fig.output-rleo}. 
Another issue might be the dust composition. 
For example, for R~Leo, \cite{cit.danchi1994} tested also a mixture of silicates and graphites. 
Yet, this led to a required stellar radius in the dust shell model 
almost twice as large as observed in the K band, 
which is supposedly sensitive to the region close to the stellar photosphere. 
Therefore, this has not been further investigated here. 
A third aspect might be the condensation temperature $T_\mathrm{cond}$ 
that we adopted. 
Some authors chose higher $T_\mathrm{cond}$, 
see, for example, \cite{cit.willson2000} for an overview. 
Taking R~Leo as study case, we set $r_\mathrm{i} = 2\,R_\star$ 
and chose $r_\mathrm{o}$ and an optical depth $\tau_{11}$ 
so that also ISI data were well fitted. 
The output of the model showed a 
somewhat increased IR excess around $11~\mathrm{\mu m}$, 
and the temperature at the inner radius 
turned out to be about $1400~\mathrm{K}$. 
Still, while the SED, K band, and ISI data were reasonably well fitted, 
the modelled visibility curve in the L band 
was clearly higher than the measured data, 
similar to Fig.~\ref{fig.output-rleo}. 

In summary, we do not believe the aspects mentioned above 
to affect our conclusions significantly.

\section{Conclusions}\label{sect.concl}
High angular resolution techniques give important constraints on
spatial intensity distribution of late-type stars.
Using diameter measurements in the K and L bands, 
we have tried to explain both K and L observations for three stars 
surrounded by circumstellar envelopes with different optical depths. 
We applied a model with a single dust shell 
and consistent with low resolution photometric data 
as well as thermal infrared interferometry data from ISI when available. 

For the supergiant Betelgeuse, 
the optically very thin dust shell has little influence 
on what is seen and is transparent in the K and L bands. 
The observed diameters are almost equal. 
In contrast, the AGB stars SW~Vir and R~Leo 
show bigger diameters in the L band than in the K band. 
Here, the optical depths of the shells are larger 
and near-infrared model visibilities react quite sensitively 
on changes of the dust shell parameters. 
Yet, the models we find in agreement with the SED, K band interferometry, and the specially dust-sensitive $11~\mathrm{\mu m}$ interferometry 
cannot reproduce the observed diameter differences. 

Our study leads to the conclusion of the insufficiency of a simple shell model 
with dust only or without more complex structures. 
New observations at several wavelengths from visible to far-infrared 
would help understand better external envelopes of AGB stars. 
A denser coverage of the spatial frequency plane 
will also be very useful to strongly constrain models. 
Interferometric observations with higher spectral resolution could 
enable to localise molecular components of these envelopes and 
certainly help explain differences between K and L band.

\appendix

\section{Evaluation of the mass-loss rate}\label{annexeA}
Let $\dot{M}_\mathrm{dust}$ be the constant mass-loss rate of dust ejected from the star, 
$m_\mathrm{d}$ the average single dust grain mass, 
and $n(r)$ the dust particle density, 
then at the inner edge $r_\mathrm{i}$ of the dust shell it holds that 
	\begin{equation}\label{equ.mass}
\dot{M}_\mathrm{dust} = 
	\frac{4\pi {r^2_\mathrm{i}} \mathrm{d}r \cdot m_\mathrm{d} n(r_\mathrm{i})}%
		{\mathrm{d}t}
	= 4\pi {r^2_\mathrm{i}} v_\mathrm{i} m_\mathrm{d} n_\mathrm{i} ,
	\end{equation}
where $v_\mathrm{i}$ is the dust outflow velocity 
and $n_\mathrm{i}=n(r_\mathrm{i})$ the particle density. 

In the dust shell, 
the particle density of grains is assumed as 
\citep{cit.schutte1989, cit.danchi1994} 
	\begin{equation}\label{equ.n_r}
n(r) = n_\mathrm{i} \cdot f(r) 
	= n_\mathrm{i} \cdot \frac{r_\mathrm{i}^2}{r^2}\cdot 
		\frac{1}{(1-A\cdot r_\mathrm{i}/r)^{1/2}}
	\end{equation}
where the parameter 
	\begin{equation}\label{equ.a_param}
A = 1-(v_\mathrm{i}/v_\mathrm{o})^2
	\end{equation}
can be adapted to different cases. 
It is zero, if $v_\mathrm{i}$ is assumed to equal 
the terminal outflow velocity $v_\mathrm{o}$ of the dust far away from the star. 
In this case, the particle density simply follows $n(r) \propto r^{-2}$. 
For taking into account acceleration of the dust by radiation pressure 
and coupling with the gas, 
$v_\mathrm{o}$ will be higher than $v_\mathrm{i}$. 
The parameter $A$ will then be larger than in the first case 
and the particle density much more pronounced at $r_\mathrm{i}$. 
In the present study we chose $A = 0.99$, 
which applies for typical conditions \citep{cit.danchi1994}. 

The optical depth $\tau_{\lambda}$ 
of the envelope is determined by 
	\begin{eqnarray}
\tau_{\lambda} & = & 
	\int_{r_\mathrm{i}}^{r_\mathrm{o}} C_\mathrm{ext}(\lambda) n(r)\,\mathrm{d}r \nonumber \\
	\label{equ.tau}
	& = & C_\mathrm{ext}(\lambda)\,n_\mathrm{i} 
		\int_{r_\mathrm{i}}^{r_\mathrm{o}} f(r)\,\mathrm{d}r ,
	\end{eqnarray}
where $n(r)$ was replaced by (\ref{equ.n_r}). 
$C_\mathrm{ext}$ is the average extinction mean section of the dust grains. 
Replacing $n_\mathrm{i}$ in (\ref{equ.mass}) by expression (\ref{equ.tau}) 
yields a relation for the mass-loss rate by dust:
	\begin{equation}\label{equ.massloss}
\dot{M}_\mathrm{dust} = 
	\frac{4\pi {r^2_\mathrm{i}} v_\mathrm{i} m_\mathrm{d} \tau_{\lambda}}%
	{C_\mathrm{ext}(\lambda)\int_{r_\mathrm{i}}^{r_\mathrm{o}}{f(r)\,\mathrm{d}r}} .
	\end{equation}

In the previous two expressions, 
$r_\mathrm{i}$, $r_\mathrm{o}$, and $\tau_\lambda$ 
are the chosen model parameters. 
$C_\mathrm{ext}$ is calculated in the modelling code 
for a given grain size distribution 
from the optical properties of the dust 
by Mie theory \citep{cit.niccolini2003}. 
Likewise, the integral in (\ref{equ.massloss}) is performed by the code. 
The dust grain mass $m_\mathrm{d}$ has to be estimated 
by averaging the size distribution $s(a)$ 
and defining the mass density. 
For the presently considered silicate dust, 
we chose as $\rho_\mathrm{d}=3.3~\mathrm{g}~\mathrm{cm}^{-3}$ 
(also used by \cite{cit.drainelee1984} and \cite{cit.danchi1994}). 
The terminal velocity of the shell particles is taken from high-spectral 
resolution observations of CO lines \citep{cit.knapp1998}. 

Furthermore, for the total mass-loss rate 
also the gas has to be taken into account, i.e., 
	\begin{equation}
\dot{M}_\mathrm{total} = \dot{M}_\mathrm{dust}+\dot{M}_\mathrm{gas} . 
	\end{equation}
Gas-to-dust mass ratios are not well known and differ quite a lot. 
We adopted here, as typical value and assuming 
dust and gas to have the same terminal velocities 
\citep{cit.schutte1989,cit.danchi1994,cit.habing1996}, 
	\begin{equation}\label{equ.g2d}
\dot{M}_\mathrm{gas} \approx 250\, \dot{M}_\mathrm{dust} .
	\end{equation}
This means that most of the mass-loss is carried by gas. 
For $\dot{M}_\mathrm{gas}$, 
an expression analogous to (\ref{equ.mass}) can be given 
where $v_\mathrm{o}$ enters for $v_\mathrm{i}$. 
Considering (\ref{equ.g2d}), 
and therefore neglecting the mass-loss carried by dust, 
the total mass-loss rate is eventually 
	\begin{equation}
\dot{M}_\mathrm{total} = 250\, 
	\frac{4\pi {r^2_\mathrm{i}} v_\mathrm{o} m_\mathrm{d} \tau_{\lambda}}%
	{C_\mathrm{ext}(\lambda)\int_{r_\mathrm{i}}^{r_\mathrm{o}}{f(r)\,\mathrm{d}r}} .
	\end{equation}

	\begin{acknowledgements}
P.~Sch.\ %
	appreciated support by the European Community 
through a Marie Curie Training Fellowship 
for an extended stay at Observatoire Paris-Meudon. 
It also allowed the collaboration with the Observatoire de la C\^{o}te d'Azur. 
-- The authors are solely responsible for the information communicated. 

This publication made use of 
NASA's Astrophysics Data System (ADS) Bibliographic Services (\url{http://cdsads.u-strasbg.fr/}) 
and of the SIMBAD database, operated at the Centre de Donn{\'e} astronomiques de Strasbourg (CDS), Strasbourg, France (\url{http://simbad.u-strasbg.fr/}). 

We thank the referee and Lee Anne Willson for helpful comments. 
	\end{acknowledgements}

\bibliographystyle{aa}   \bibliography{lit-arts,lit-procs,lit-dra-upd,lit-base}   %

\begin{thebibliography}{36}
\expandafter\ifx\csname natexlab\endcsname\relax\def\natexlab#1{#1}\fi

\bibitem[{ADS(2003)}]{ads}
ADS. 2003, {Astrophysics Data System}, \url{http://cdsads.u-strasbg.fr/}

\bibitem[{{Benson} {et~al.}(1989){Benson}, {Turner}, \&
  {Dyck}}]{cit.benson1989}
{Benson}, J.~A., {Turner}, N.~H., \& {Dyck}, H.~M. 1989, \aj, 97, 1763

\bibitem[{{Chagnon} {et~al.}(2002){Chagnon}, {Mennesson}, {Perrin}, {Coud{\'e}
  du Foresto}, {Salom{\'e}}, {Bord{\'e}}, {Lacasse}, \&
  {Traub}}]{cit.chagnon2002}
{Chagnon}, G., {Mennesson}, B., {Perrin}, G., {et~al.} 2002, \aj, 124, 2821

\bibitem[{{Coud{\'e} du Foresto} {et~al.}(1998){Coud{\'e} du Foresto},
  {Perrin}, {Ruilier}, {Mennesson}, {Traub}, \& {Lacasse}}]{cit.foresto1998}
{Coud{\'e} du Foresto}, V., {Perrin}, G., {Ruilier}, C., {et~al.} 1998, in SPIE
  Proceedings Series, Vol. 3350, Astronomical Interferometry, ed. R.~D.
  {Reasenberg}, 856--863

\bibitem[{{Danchi} {et~al.}(1994){Danchi}, {Bester}, {Degiacomi}, {Greenhill},
  \& {Townes}}]{cit.danchi1994}
{Danchi}, W.~C., {Bester}, M., {Degiacomi}, C.~G., {Greenhill}, L.~J., \&
  {Townes}, C.~H. 1994, \aj, 107, 1469

\bibitem[{{Draine} \& {Lee}(1984)}]{cit.drainelee1984}
{Draine}, B.~T. \& {Lee}, H.~M. 1984, \apj, 285, 89

\bibitem[{{Gail} \& {Sedlmayr}(1998)}]{cit.gail1998}
{Gail}, H.-P. \& {Sedlmayr}, E. 1998, in The Molecular Astrophysics of Stars
  and Galaxies, ed. T.~W. {Hartquist} \& D.~A. {Williams} (Oxford University
  Press), 285--312

\bibitem[{{Habing}(1996)}]{cit.habing1996}
{Habing}, H.~J. 1996, \aapr, 7, 97

\bibitem[{{Hale} {et~al.}(2000){Hale}, {Bester}, {Danchi}, {Fitelson}, {Hoss},
  {Lipman}, {Monnier}, {Tuthill}, \& {Townes}}]{cit.ISI}
{Hale}, D.~D.~S., {Bester}, M., {Danchi}, W.~C., {et~al.} 2000, \apj, 537, 998

\bibitem[{{Kholopov} {et~al.}(1998){Kholopov}, {Samus}, {Frolov}, {Goranskij},
  {Gorynya}, {Karitskaya}, {Kazarovets}, {Kireeva}, {Kukarkina}, {Kurochkin},
  {Medvedeva}, {Pastukhova}, {Perova}, {Rastorguev}, \&
  {Shugarov}}]{cit.gcvs1998}
{Kholopov}, P.~N., {Samus}, N.~N., {Frolov}, M.~S., {et~al.} 1998, VizieR
  Online Data Catalogue, {Combined General Catalogue of Variable Stars,
  \url{http://cdsweb.u-strasbg.fr/viz-bin/VizieR?-source=II/214}}

\bibitem[{{Knapp} {et~al.}(1998){Knapp}, {Young}, {Lee}, \&
  {Jorissen}}]{cit.knapp1998}
{Knapp}, G.~R., {Young}, K., {Lee}, E., \& {Jorissen}, A. 1998, \apjs, 117, 209

\bibitem[{{Lafon} \& {Berruyer}(1991)}]{cit.lafon1991}
{Lafon}, J.-P.~J. \& {Berruyer}, N. 1991, \aapr, 2, 249

\bibitem[{{Lopez} {et~al.}(1997)}]{cit.lopez1997}
{Lopez}, B. {et~al.} 1997, \apj, 488, 807

\bibitem[{{Mathis} {et~al.}(1977){Mathis}, {Rumpl}, \&
  {Nordsieck}}]{cit.mathis1977}
{Mathis}, J.~S., {Rumpl}, W., \& {Nordsieck}, K.~H. 1977, \apj, 217, 425

\bibitem[{{Mennesson} {et~al.}(1999){Mennesson}, {Mariotti}, {Coud{\'e} du
  Foresto}, {Perrin}, {Ridgway}, {Ruilier}, {Traub}, {Carleton}, {Lacasse}, \&
  {Maz{\'e}}}]{cit.mennesson1999}
{Mennesson}, B., {Mariotti}, J.~M., {Coud{\'e} du Foresto}, V., {et~al.} 1999,
  \aap, 346, 181

\bibitem[{{Mennesson} {et~al.}(2002){Mennesson}, {Perrin}, {Chagnon},
  {Foresto}, {Ridgway}, {Merand}, {Salome}, {Borde}, {Cotton}, {Morel},
  {Kervella}, {Traub}, \& {Lacasse}}]{cit.mennesson2002}
{Mennesson}, B., {Perrin}, G., {Chagnon}, G., {et~al.} 2002, \apj, 579, 446

\bibitem[{{Monnier} {et~al.}(1998){Monnier}, {Geballe}, \&
  {Danchi}}]{cit.monnier1998}
{Monnier}, J.~D., {Geballe}, T.~R., \& {Danchi}, W.~C. 1998, \apj, 502, 833

\bibitem[{{Niccolini} {et~al.}(2003){Niccolini}, {Woitke}, \&
  {Lopez}}]{cit.niccolini2003}
{Niccolini}, G., {Woitke}, P., \& {Lopez}, B. 2003, \aap, 399, 703

\bibitem[{{Perrin}(2003)}]{cit.perrin2003a}
{Perrin}, G. 2003, \aap, 400, 1173

\bibitem[{{Perrin} {et~al.}(1998){Perrin}, {Coud{\'e} du Foresto}, {Ridgway},
  {Mariotti}, {Traub}, {Carleton}, \& {Lacasse}}]{cit.teff}
{Perrin}, G., {Coud{\'e} du Foresto}, V., {Ridgway}, S.~T., {et~al.} 1998,
  \aap, 331, 619

\bibitem[{{Perrin} {et~al.}(1999){Perrin}, {Coud{\'e} du Foresto}, {Ridgway},
  {Mennesson}, {Ruilier}, {Mariotti}, {Traub}, \& {Lacasse}}]{cit.perrin1999}
{Perrin}, G., {Coud{\'e} du Foresto}, V., {Ridgway}, S.~T., {et~al.} 1999,
  \aap, 345, 221

\bibitem[{{Perrin} {et~al.}(2003{\natexlab{a}})}]{cit.perrin2002aori}
{Perrin}, G. {et~al.} 2003{\natexlab{a}}, \aap, submitted

\bibitem[{{Perrin} {et~al.}(2003{\natexlab{b}})}]{cit.perrin2003c}
{Perrin}, G. {et~al.} 2003{\natexlab{b}}, in preparation

\bibitem[{{Perryman} {et~al.}(1997){Perryman}, {Lindegren}, {Kovalevsky},
  {Hoeg}, {Bastian}, {Bernacca}, {Cr{\'e}z{\'e}}, {Donati}, {Grenon}, {van
  Leeuwen}, {van der Marel}, {Mignard}, {Murray}, {Le Poole}, {Schrijver},
  {Turon}, {Arenou}, {Froeschl{\'e}}, \& {Petersen}}]{cit.hip1997}
{Perryman}, M.~A.~C., {Lindegren}, L., {Kovalevsky}, J., {et~al.} 1997, \aap,
  323, L49, {VizieR Online Data Catalogue, HIPPARCOS Catalogue,
  \url{http://cdsweb.u-strasbg.fr/viz-bin/VizieR?-source=I/239}}

\bibitem[{{Schutte} \& {Tielens}(1989)}]{cit.schutte1989}
{Schutte}, W.~A. \& {Tielens}, A.~G.~G.~M. 1989, \apj, 343, 369

\bibitem[{SIMBAD(2003)}]{simbad}
SIMBAD. 2003, {SIMBAD Astronomical Database}, \url{http://simbad.u-strasbg.fr/}

\bibitem[{{Strecker} {et~al.}(1978){Strecker}, {Erickson}, \&
  {Witteborn}}]{cit.strecker1978}
{Strecker}, D.~W., {Erickson}, E.~F., \& {Witteborn}, F.~C. 1978, \aj, 83, 26

\bibitem[{{Suh}(1999)}]{cit.suh1999}
{Suh}, K. 1999, \mnras, 304, 389

\bibitem[{{Traub}(1998)}]{cit.traub1998}
{Traub}, W.~A. 1998, in SPIE Proceedings Series, Vol. 3350, Astronomical
  Interferometry, ed. R.~D. {Reasenberg}, 848--855

\bibitem[{{Tsuji}(2000)}]{cit.tsuji2000}
{Tsuji}, T. 2000, \apj, 538, 801

\bibitem[{{Tsuji} {et~al.}(1997){Tsuji}, {Ohnaka}, {Aoki}, \&
  {Yamamura}}]{cit.tsuji1997}
{Tsuji}, T., {Ohnaka}, K., {Aoki}, W., \& {Yamamura}, I. 1997, \aap, 320, L1

\bibitem[{{van der Veen} {et~al.}(1995){van der Veen}, {Omont}, {Habing}, \&
  {Matthews}}]{cit.veen1995}
{van der Veen}, W.~E.~C.~J., {Omont}, A., {Habing}, H.~J., \& {Matthews}, H.~E.
  1995, \aap, 295, 445

\bibitem[{{Wannier} \& {Sahai}(1986)}]{cit.wannier1986}
{Wannier}, P.~G. \& {Sahai}, R. 1986, \apj, 311, 335

\bibitem[{{Weiner} {et~al.}(2000)}]{cit.weiner2000}
{Weiner}, J. {et~al.} 2000, Astrophys.\ J., 544, 1097

\bibitem[{{Willson}(2000)}]{cit.willson2000}
{Willson}, L.~A. 2000, Ann.\ Rev.\ Astron.\ Astrophys., 38, 573

\bibitem[{{Yamamura} {et~al.}(1999){Yamamura}, {de Jong}, \&
  {Cami}}]{cit.yamamura1999}
{Yamamura}, I., {de Jong}, T., \& {Cami}, J. 1999, \aap, 348, L55

\end{thebibliography}

\end{document}